\documentclass{lmcs}
\pdfoutput=1
\usepackage[utf8]{inputenc}

\usepackage{lastpage}
\lmcsdoi{19}{1}{9}
\lmcsheading{}{\pageref{LastPage}}{}{}%
{Dec.~30,~2021}{Feb.~01,~2023}{}

\usepackage{hyperref}
\usepackage[english]{babel}
\usepackage{algorithm}
\usepackage[noend]{algorithmic}
\usepackage[all]{hypcap} 
\usepackage{amssymb}

\begin{document}

\title[A proof system for graph (non)-isomorphism verification]{A proof system for graph (non)-isomorphism verification}
\author[M.~Bankovi\' c]{Milan Bankovi\' c\lmcsorcid{0000-0002-0517-6334}}
\address{Faculty of Mathematics, University of Belgrade, Studentski Trg 16, 11000 Belgrade, Serbia}
\email{milan@matf.bg.ac.rs}

\author[I.~Drecun]{Ivan Drecun\lmcsorcid{0000-0001-7145-3662}}
\email{ivan\_drecun@matf.bg.ac.rs}

\author[F.~Mari\' c]{Filip Mari\' c\lmcsorcid{0000-0001-7219-6960}}
\email{filip@matf.bg.ac.rs}

\begin{abstract}
  In order to apply canonical labelling of graphs and isomorphism
  checking in interactive theorem provers, these checking algorithms
  must either be mechanically verified or their results must be
  verifiable by independent checkers. We analyze a state-of-the-art
  algorithm for canonical labelling of graphs (described by McKay and
  Piperno) and formulate it in terms of a formal proof system. We
  provide an implementation that can export a proof that the obtained
  graph is the canonical form of a given graph. Such proofs are then
  verified by our independent checker and can be used to confirm that
  two given graphs are not isomorphic.
\end{abstract}

\maketitle

\section{Introduction}

An isomorphism between two graphs is a bijection between their vertex
sets that preserves adjacency. Testing whether two given graphs are
isomorphic is a problem that has been studied extensively since the
1970s (so much so that it has even been called a ``disease'' among
algorithm designers). Since then many efficient algorithms have been
proposed and practically used
(e.g.,~\cite{babai,mckay_practical,mckay_piperno,conauto}). Most
state-of-the-art algorithms and tools for isomorphism checking
(e.g.,~\verb|nauty|, \verb|bliss|, \verb|saucy|, and \verb|Traces|)
are based on \emph{canonical labellings} \cite{mckay_piperno}. They
assign a unique vertex labelling to each graph, so that two graphs are
isomorphic if and only if their canonical labellings are the
same. Such algorithms can also compute automorphism groups of
graphs. Graph isomorphism checking and the computation of canonical
labellings and automorphism groups are useful in many
applications. For example, they are used in many algorithms that
enumerate and count isomorph-free families of some combinatorial
objects \cite{mckay_isomorph_free}. Traditional applications of graph
isomorphism testing are found in mathematical chemistry
(e.g.~identifying chemical compounds in chemical databases) and in
electronic design automation (e.g., checking whether electronic
circuits represented by a schematic and an integrated circuit layout
are identical), and there are more recent applications in machine
learning (e.g.,~\cite{pmlr-v48-niepert16}).

In the past decades, we have witnessed the rise of automated and
interactive theorem proving and its application in formalizing many
areas of mathematics and computer science
\cite{ams-formal,itp_milestones}.  Some of the most famous results in
interactive theorem proving had significant combinatorial
components. For example, the proof of the four-color theorem
\cite{four-colour} required the analysis of several hundred
non-isomorphic graph configurations, in some cases involving more than
20 million different cases. The formal proof of Kepler's conjecture
about the optimal packing of spheres in three-dimensional Euclidean
space \cite{flyspeck} required the identification, enumeration and
analysis of several thousand non-isomorphic graphs \cite{tame-graphs},
which was done using a custom formally verified algorithm. Initial
steps to verify generic algorithms for the isomorph-free enumeration
of combinatorial objects have been taken (e.g., the Farad\v zev-Read
enumeration has been formally verified within Isabelle/HOL
\cite{faradzev-read}). Several other such algorithms (e.g.,
\cite{mckay_isomorph_free}) would benefit from trustworthy graph
isomorphism checking, automorphism group computation, and canonical
labelling.

Isomorphism checking can be used in theorem proving applications only
if it is somehow verified. The simple case is when an isomorphism
between two graphs is explicitly constructed, since it is then easy to
independently verify that the graphs are isomorphic. However, if an
algorithm asserts that the given graphs are not isomorphic, that
assertion is not easily independently verifiable. One approach would
be to fully formally verify the isomorphism checking algorithm and its
implementation in an interactive theorem prover. This would be a very
difficult and tedious task. The other approach, which we are
investigating in this paper, is to extend the isomorphism checking
algorithm so that it can generate \emph{certificates} for
non-isomorphism. These certificates would then be verified by an
independent checker, which is much simpler than the isomorphism
checking algorithm itself. A similar approach has already been
successfully used in SAT and SMT solving
\cite{drat-trim,largest-proof,lammich_unsat}. In particular, we will
extend a canonical labelling algorithm to produce a certificate
confirming that the canonical labelling has been correctly computed.

In the rest of the paper we will examine the scheme for canonical
labelling and graph isomorphism checking proposed by McKay and Piperno
\cite{mckay_piperno} and formulate it in the form of a proof system
(in the spirit of the proof systems for SAT and SMT
\cite{KrsticSMT,DPLLT,MJSAT}). We prove the soundness and completeness
of the proof system and implement a proof checker for it. We develop
our own implementation of McKay and Piperno's algorithm, which exports
canonical labelling certificates (proofs in our proof-system). These
certificates are then checked independently by our proof checker. We
also report experimental results. Finally, we formalize central
results (the correctness of McKay/Piperno's abstract scheme and the
soundness of our proof system) in Isabelle/HOL\footnote{The whole
  formalization is available at
  \url{https://github.com/milanbankovic/isocert}}.

\section{Background}

Notation used in this text is mostly borrowed from McKay and Piperno
\cite{mckay_piperno}.
Let $G = (V, E)$ be an undirected graph with the set of vertices
$V = \{ 1, 2,\ldots, n \}$ and the set of edges
$E \subseteq V \times V$. Let $\pi$ be a \emph{coloring} of the
graph's vertices, i.e.~a surjective function from $V$ to
$\{ 1, \ldots, m \}$. The set $\pi^{-1}(k)$ ($1 \le k \le m$) is a
\emph{cell} of a coloring $\pi$ corresponding to the color $k$
(i.e.~the set of vertices from $G$ colored with the color $k$). We
will also call this cell \emph{the $k$-th cell of $\pi$}, and the
coloring $\pi$ can be represented by the sequence
$\pi^{-1}(1),\pi^{-1}(2),\ldots,\pi^{-1}(m)$ of its cells. The pair
$(G, \pi)$ is called a \emph{colored graph}.

A coloring $\pi'$ is \emph{finer} than $\pi$ (denoted by
$\pi' \preceq \pi$) if $\pi(u) < \pi(v)$ implies $\pi'(u) < \pi'(v)$
for all $u,v \in V$ (note that this implies that each cell of $\pi'$
is a subset of some cell of $\pi$). A coloring $\pi'$ is
\emph{strictly finer} than $\pi$ (denoted by $\pi' \prec \pi$) if
$\pi' \preceq \pi$ and $\pi' \ne \pi$. A coloring $\pi$ is
\emph{discrete} if all its cells are singleton. In that case, the
coloring $\pi$ is a permutation of $V$ (i.e.~the colors of $\pi$ can
be considered as unique labels of the graph's vertices).

A permutation $\sigma \in S_n$ acts on a graph $G = (V, E)$ by
permuting the labels of vertices, i.e.~the resulting graph is
$G^\sigma = (V^\sigma, E^\sigma)$, where $V^\sigma = V$, and
$E^\sigma = \{ (u^\sigma, v^\sigma)\ |\ (u,v) \in E\}$. The
permutation $\sigma$ acts on a coloring $\pi$ such that for the
resulting coloring $\pi^\sigma$ it holds that
$\pi^\sigma(v^\sigma) = \pi(v)$ (that is, the $\sigma$-images of
vertices are colored by $\pi^\sigma$ in the same way as their
originals were colored by $\pi$). For a colored graph $(G,\pi)$, we
define $(G, \pi)^\sigma = (G^\sigma, \pi^\sigma)$.

The composition of permutations $\sigma_1$ and $\sigma_2$ is denoted
by $\sigma_1\sigma_2$. It holds
$(G, \pi)^{\sigma_1\sigma_2} = ((G,\pi)^{\sigma_1})^{\sigma_2} =  (G^{\sigma_1},
\pi^{\sigma_1})^{\sigma_2} = ((G^{\sigma_1})^{\sigma_2},
(\pi^{\sigma_1})^{\sigma_2})$. It can be noticed that, if a discrete
coloring $\pi$ is seen as a permutation, it holds
$\pi^\sigma = \sigma^{-1}\pi$, for each $\sigma \in S_n$.

Two graphs $G_1$ and $G_2$ (or colored graphs $(G_1, \pi_1)$ and
$(G_2,\pi_2)$) are \emph{isomorphic} if there exists a permutation
$\sigma \in S_n$, such that $G_1^\sigma = G_2$ (or
$(G_1^\sigma, \pi_1^\sigma) = (G_2,\pi_2)$, respectively). 
The problem of \emph{graph isomorphism} is the problem of determining
whether two given (colored) graphs are isomorphic.

A permutation $\sigma \in S_n$ is an \emph{automorphism} of a graph
$G$ (or a colored graph $(G,\pi)$), if $G^\sigma = G$ (or
$(G^\sigma, \pi^\sigma) = (G, \pi)$, respectively). The set of all
automorphisms of a colored graph $(G,\pi)$ is denoted by
$Aut(G,\pi)$. It is a subgroup of $S_n$.
The \emph{orbit} of a vertex $v \in V$ with respect to $Aut(G,\pi)$ 
is the set $\Omega(v,G,\pi) = \{ v^\sigma \ |\ \sigma \in Aut(G,\pi) \}$.

\section{McKay and Piperno's scheme}
\label{sec:mckay}

The graph isomorphism algorithm considered in this text is described
by McKay and Piperno in \cite{mckay_piperno}. The central idea is to
construct the \emph{canonical form} of a graph $(G,\pi)$ (denoted by
$C(G,\pi)$) which is a colored graph obtained by relabelling the
vertices of $G$ (i.e.~there is a permutation $\sigma \in S_n$ such
that $C(G,\pi)=(G,\pi)^\sigma$), such that the canonical forms of two
graphs are identical if and only if the graphs are isomorphic. In this
way, the isomorphism of two given graphs can be checked by computing
canonical forms of both graphs and then comparing them for equality.

McKay and Piperno describe an abstract framework for computing
canonical forms. It is parameterised by several abstract functions
that, when fixed, yield a concrete, deterministic algorithm for
computing canonical forms. Any choice of these functions that
satisfies certain properties (given as axioms) will result in a
correct algorithm.

In the abstract framework, the canonical form of a graph is computed
by constructing the \emph{search tree} of a graph $(G,\pi_0)$, where
$\pi_0$ is the initial coloring of the graph $G$. This tree is
denoted by $\mathcal{T}(G, \pi_0)$. Each node of this tree is
associated with a unique sequence $\nu = [v_1,v_2,\ldots,v_k]$ of
distinct vertices of $G$. The root of the tree is assigned the empty
sequence $[\ ]$. The length of a sequence $\nu$ is denoted by $|\nu|$,
and the prefix of a length $i$ of a sequence $\nu$ is denoted by
$[\nu]_i$. The subtree of $\mathcal{T}(G,\pi_0)$ rooted at a node
$\nu$ is denoted by $\mathcal{T}(G,\pi_0,\nu)$. A permutation
$\sigma \in S_n$ acts on search (sub)trees by acting on each of the
sequences associated with its nodes.

Each node $\nu$ of the search tree is also assigned a coloring of $G$,
denoted by $R(G,\pi_0,\nu)$. The \emph{refinement function} $R$ is one
of the parameters that must satisfy the following three properties for the
algorithm to be correct:

\begin{enumerate}[align=left]
\item[(R1)] The coloring $R(G, \pi_0, \nu)$ must be finer than $\pi_0$.
\item[(R2)] It must individualize the vertices from $\nu$, i.e.~each vertex
  from $\nu$ must be in a singleton cell of $R(G,\pi_0,\nu)$.
\item[(R3)] The function $R$ must be \emph{label invariant}, that is, it
  must hold that
  $R(G,\pi_0,\nu)^\sigma = R(G^\sigma, \pi_0^\sigma, \nu^\sigma)$.
\end{enumerate}

If the coloring assigned to a node $\nu$ is discrete, then this node
is a leaf of the search tree. Otherwise, we choose some non-singleton
cell of the coloring $R(G,\pi_0,\nu)$, which we call the \emph{target
  cell} and denote by $T(G, \pi_0, \nu)$. The function $T$ is called
\emph{target cell selector} and is another parameter that must satisfy
the following properties for the algorithm to be correct:

\begin{enumerate}[align=left]
\item[(T1)] For leaves of the search tree $T$ returns an empty cell.
\item[(T2)] For inner nodes of the search tree $T$ returns some
  non-singleton cell of $R(G,\pi_0,\nu)$.
\item[(T3)] The function $T$ must be label invariant, i.e.~it must hold
  $T(G,\pi_0,\nu)^\sigma = T(G^\sigma, \pi_0^\sigma, \nu^\sigma)$.
\end{enumerate}

If $T(G,\pi_0,\nu) = \{ w_1, \ldots, w_m \}$, then the children of the
node $\nu = [v_1,\ldots,v_k]$ in $\mathcal{T}(G,\pi_0)$ are the nodes
$[\nu,w_1]$, $[\nu,w_2]$,\ldots, $[\nu,w_m]$ in this order, where
$[\nu,w_i]$ denotes the sequence $[v_1,\ldots,v_k,w_i]$. Note that the
coloring of each child of the node $\nu$ will individualize at least
one more vertex, which guarantees that the tree $\mathcal{T}(G,\pi_0)$
will be finite. Note also that if the order of the child nodes is
fixed (and this can be achieved by assuming that the cell elements
$w_1$ to $w_m$ are given in ascending order), once the functions $R$
and $T$ are fixed, the tree $\mathcal{T}(G, \pi_0)$ is uniquely
determined from the original colored graph $(G, \pi_0)$.

Since the coloring $\pi = R(G,\pi_0,\nu)$ is discrete if and only if
the node $\nu$ is a leaf, and since discrete colorings can be viewed
as permutations of the graph's vertices, a graph $G_\nu = G^\pi$ can
be associated to any leaf $\nu$.

\begin{exa}
  In the example in Figure~\ref{fig:tree}, the target cell is always
  the first non-singleton cell of $R(G, \pi_0, \nu)$ (in each internal
  search tree node there is only one non-singleton cell, therefore the
  target cell selector $T$ is trivial). The coloring
  $R(G, \pi_0, [\ ])$ is equal to $\pi_0$, and
  $\pi = R(G, \pi_0, [\nu, w])$ introduces new cells with respect to
  $R(G, \pi_0, \nu)$ only by individualizing $w$. The resulting
  colorings are shown under each node in Figure~\ref{fig:tree} (as
  sequences of cells, that are sets of vertices).

\begin{figure}[!ht]
\begin{center}
 \begin{tikzpicture}
  \begin{scope}
    \node (1) at (0, 0)[draw, circle, label=180:$a$] {1};
    \node (2) at (0.5, 0.86)[draw, circle, label=90:$b$] {1};
    \node (3) at (1, 0)[draw, circle, label=0:$c$] {1};
  \end{scope}
  \begin{scope}
    \path [-] (1) edge (2);
    \path [-] (2) edge (3);
  \end{scope}
\end{tikzpicture}

\begin{tikzpicture}
  \begin{scope}
    \node (0) at (0, 1.5) {$\nu=[\,]$, $\{a,b,c\}$};
    
    \node (1a) at (-3.2, 0.2) {};
    \node (2a) at (0, 0) {};
    \node (3a) at (3.2, 0.2) {};
    
    \node (1) at (-4, -1) {$\nu=[a]$, $\{a\}\{b,c\}$};
    \node (2) at (0, -1) {$\nu=[b]$, $\{b\}\{a,c\}$};
    \node (3) at (4, -1) {$\nu=[c]$, $\{c\}\{a,b\}$};
    \node (12) at (-5, -1.7) {};
    \node (13) at (-3, -1.7) {};
    \node (21) at (-1, -1.7) {};
    \node (23) at (1, -1.7) {};
    \node (31) at (3, -1.7) {};
    \node (32) at (5, -1.7) {};
    
    \node (12a) at (-5, -4) {$\{a\}\{b\}\{c\}$};
    \node (13a) at (-3, -4) {$\{a\}\{c\}\{b\}$};
    \node (21a) at (-1, -4) {$\{b\}\{a\}\{c\}$};
    \node (23a) at (1, -4) {$\{b\}\{c\}\{a\}$};
    \node (31a) at (3, -4) {$\{c\}\{a\}\{b\}$};
    \node (32a) at (5, -4) {$\{c\}\{b\}\{a\}$};

    \node (12b) at (-5, -3.5) {$\nu = [a, b]$,};
    \node (13b) at (-3, -3.5) {$\nu = [a, c]$,};
    \node (21b) at (-1, -3.5) {$\nu = [b, a]$,};
    \node (23b) at (1, -3.5) {$\nu = [b, c]$,};
    \node (31b) at (3, -3.5) {$\nu = [c, a]$,};
    \node (32b) at (5, -3.5) {$\nu = [c, b]$,};

    \node (a1) at (-5.5, -2.86)[draw, circle] {1};
    \node (a2) at (-5, -2)[draw, circle] {2};
    \node (a3) at (-4.5, -2.86)[draw, circle] {3};

    \node (b1) at (-3.5, -2.86)[draw, circle] {1};
    \node (b2) at (-3, -2)[draw, circle] {3};
    \node (b3) at (-2.5, -2.86)[draw, circle] {2};

    \node (c1) at (-1.5, -2.86)[draw, circle] {2};
    \node (c2) at (-1, -2)[draw, circle] {1};
    \node (c3) at (-0.5, -2.86)[draw, circle] {3};

    \node (d1) at (0.5, -2.86)[draw, circle] {3};
    \node (d2) at (1, -2)[draw, circle] {1};
    \node (d3) at (1.5, -2.86)[draw, circle] {2};

    \node (e1) at (2.5, -2.86)[draw, circle] {2};
    \node (e2) at (3, -2)[draw, circle] {3};
    \node (e3) at (3.5, -2.86)[draw, circle] {1};

    \node (f1) at (4.5, -2.86)[draw, circle] {3};
    \node (f2) at (5, -2)[draw, circle] {2};
    \node (f3) at (5.5, -2.86)[draw, circle] {1};

    \node (g1) at (-4.5, -0.26)[draw, circle] {1};
    \node (g2) at (-4, 0.6)[draw, circle] {2};
    \node (g3) at (-3.5, -0.26)[draw, circle] {2};

    \node (h1) at (-0.5, -0.26)[draw, circle] {2};
    \node (h2) at (0, 0.6)[draw, circle] {1};
    \node (h3) at (0.5, -0.26)[draw, circle] {2};

    \node (i1) at (4.5, -0.26)[draw, circle] {1};
    \node (i2) at (4, 0.6)[draw, circle] {2};
    \node (i3) at (3.5, -0.26)[draw, circle] {2};

    \node (dummy1) at (0, 0.85)[opacity=0] {};
    \node (dummy2) at (0, -0.3)[opacity=0] {};
  \end{scope}
  \begin{scope}
    \path [-] (0) edge (1a);
    \path [-] (0) edge (dummy1);
    \path [-] (0) edge (3a);
    \path [-] (1) edge (12);
    \path [-] (1) edge (13);
    \path [-] (2) edge (21);
    \path [-] (2) edge (23);
    \path [-] (3) edge (31);
    \path [-] (3) edge (32);

    \path [-] (a1) edge (a2);
    \path [-] (a2) edge (a3);

    \path [-] (b1) edge (b2);
    \path [-] (b2) edge (b3);

    \path [-] (c1) edge (c2);
    \path [-] (c2) edge (c3);

    \path [-] (d1) edge (d2);
    \path [-] (d2) edge (d3);

    \path [-] (e1) edge (e2);
    \path [-] (e2) edge (e3);

    \path [-] (f1) edge (f2);
    \path [-] (f2) edge (f3);

    \path [-] (g1) edge (g2);
    \path [-] (g2) edge (g3);

    \path [-] (h1) edge (h2);
    \path [-] (h2) edge (h3);

    \path [-] (i1) edge (i2);
    \path [-] (i2) edge (i3);
  \end{scope}
\end{tikzpicture}
\end{center}
\caption{A search tree that generates all permutations of the vertex
  set. In each search tree node, the colored graph (each graph node
  contains the number denoting its color), the sequence $\nu$, and the
  coloring $R(G, \pi_0, \nu)$ (given as a sequence of its cells) are
  shown.}
\label{fig:tree}
\end{figure} 
\end{exa}

Each node $\nu$ of $\mathcal{T}(G,\pi_0)$ is also assigned a
\emph{node invariant}, denoted by $\phi(G,\pi_0,\nu)$. Node invariants
map tree nodes to elements of some totally ordered set (usually
lexicographically ordered sequences of numbers).

The node invariant function $\phi$ is another parameter of the
algorithm, and must satisfy the following conditions for the algorithm
to be correct:

\begin{enumerate}[align=left]
\item[($\phi$1)] If $|\nu'| = |\nu''|$ and
  $\phi(G,\pi_0,\nu') < \phi(G,\pi_0,\nu'')$, then for every leaf
  $\nu_1 \in \mathcal{T}(G, \pi_0, \nu')$ and leaf
  $\nu_2 \in \mathcal{T}(G, \pi_0, \nu'')$ it holds
  $\phi(G, \pi_0, \nu_1) < \phi(G, \phi_0, \nu_2)$
\item[($\phi$2)] If $\pi' = R(G, \pi_0, \nu')$ and
  $\pi'' = R(G, \pi_0, \nu'')$ are discrete, and if
  $\phi(G, \pi_0, \nu') = \phi(G, \pi_0, \nu'')$, then
  $G^{\pi'} = G^{\pi''}$ (the last relation is equality, not
  isomorphism)
\item[($\phi$3)]
  $\phi(G^\sigma, \pi_0^\sigma, \nu^\sigma) = \phi(G,\pi_0,\nu)$ for
  each node $\nu$ and for each permutation $\sigma$ (that is, $\phi$
  is label invariant)
\end{enumerate}

Let
$\phi(G, \pi_0, \nu) = [f(G, \pi_0, [\nu]_0), f(G, \pi_0, [\nu]_1),
\dots, f(G, \pi_0, [\nu]_{|\nu|})]$ for some label invariant function
$f$ that maps the tree nodes to elements of a totally ordered set.
Then $\phi$ trivially
satisfies the axioms $\phi1$ and $\phi3$, assuming the
lexicographical order.
To
ensure that the axiom $\phi$2 also holds, for each leaf node $\nu$ we must append
$g(G_{\nu})$ to this list (i.e.~for leaves we define $\phi(G,\pi_0,\nu)= [f(G, \pi_0, [\nu]_0), f(G, \pi_0, [\nu]_1),
\dots, f(G, \pi_0, [\nu]_{|\nu|}), g(G_{\nu})]$), where $g$ is some injective function that encodes a
graph by an element of some totally ordered set (e.g., it could be some
representation of the graph's adjacency matrix).

To simplify the formal setup and avoid the need of appending the graph
representation $g(G_{\nu})$ to the list of values of the function $f$,
we can slightly change the first two axioms for $\phi$ (the axiom
$\phi3$ remains unchanged):

\begin{enumerate}[align=left]
\item[($\phi$1)] If $|\nu'| = |\nu''|$ and $\phi(G,\pi_0,\nu') < \phi(G,\pi_0,\nu'')$,
  then
  $\phi(G,\pi_0,[\nu', w_1]) <
  \phi(G,\pi_0,[\nu'',w_2])$ for each
  $w_1 \in T(G,\pi_0,\nu')$,
  $w_2 \in T(G,\pi_0,\nu'')$.
\item[($\phi$2)] For each $\nu$ and nonempty $\nu'$ such that
  $[\nu, \nu']$ is a tree node, it holds that
  $\phi(G, \pi_0, \nu) < \phi(G, \pi_0, [\nu, \nu'])$.
  \item[($\phi$3)]
  $\phi(G^\sigma, \pi_0^\sigma, \nu^\sigma) = \phi(G,\pi_0,\nu)$ for
  each node $\nu$ and for each permutation $\sigma$.
\end{enumerate}
In this setup, we only require that $\phi$ behaves like a lexicographic order,
and omit graph comparison present in the original axiom $\phi$2. Now, we can define the node
invariant function $\phi$ uniformly, for all nodes $\nu$,
as
$\phi(G, \pi_0, \nu) = [f(G, \pi_0, [\nu]_0), f(G, \pi_0, [\nu]_1),
\dots, f(G, \pi_0, [\nu]_{|\nu|})]$, and the new axioms trivially hold.

Let
$\phi_{max} = \max \{ \phi(G,\pi_0,\nu)\ |\ \nu\ \text{is a leaf}
\}$. A \emph{maximal leaf} is any leaf $\nu$ such that
$\phi(G,\pi_0,\nu) = \phi_{max}$. Notice that with our new axioms
there can be multiple maximal leaves and, contrary to the original axioms,
their corresponding graphs may
be distinct. Let us assume that there is a total ordering defined on
graphs (for instance, we can compare their adjacency matrices
lexicographically). Let
$G_{max} = \max\{ G_\nu\ |\ \nu\ \text{is a maximal leaf} \}$ (that
is, $G_{max}$ is a maximal graph assigned to some maximal
leaf). Again, there may be more than one maximal leaves corresponding
to $G_{max}$. Among them, the one
labeled with the lexicographically smallest sequence will be called the
\emph{canonical leaf} of the search tree $\mathcal{T}(G,\pi_0)$, and will be denoted
by $\nu^*$. Let $\pi^* = R(G,\pi_0,\nu^*)$ be the coloring of the leaf
$\nu^*$. The graph $G^{\pi^*} = G_{max}$ that corresponds to the
canonical leaf $\nu^*$ is the canonical form of $(G,\pi_0)$, that is,
we define $C(G,\pi_0) = (G^{\pi^*}, \pi_0^{\pi^*})$. So, in order to
find the canonical form of the graph, we must find the canonical leaf of
the search tree.

Note that once the node invariant function $\phi$, and the graph
ordering are fixed, the canonical leaf $\nu^*$ and the canonical form
$C(G, \pi_0)$ are uniquely
defined based only on the initial colored graph $(G, \pi_0)$.

\begin{exa}
  Continuing the previous example, let $f(G, \pi_0, \nu)$ be the
  maximal vertex degree of the cell $R(G, \pi_0, \nu)^{-1}(1)$ and let
  $\phi(G, \pi_0, \nu)$ be defined as the sequence
  $[f(G, \pi_0, [\nu]_0), f(G, \pi_0, [\nu]_1), \dots, f(G, \pi_0,
  [\nu]_{|\nu|})]$.  Note that $f$ is label invariant, implying that
  $\phi$ is a valid node invariant. The node invariants for the nodes
  of the search tree from Figure~\ref{fig:tree}, the adjacency
  matrices of the graphs corresponding to the leaves of the tree, and
  the canonical leaf of the tree are shown in Figure~\ref{fig:invariant}.

\begin{figure}[h!]
\begin{center}
 \begin{tikzpicture}
  \begin{scope}
    \node (0) at (0, 0) [label=0:{[2]}] {$\{a,b,c\}$};
    \node (1) at (-4, -1) [label=0:{[2,1]}] {$\{a\}\{b,c\}$};
    \node (2) at (0, -1) [label=0:{[2,2]}] {$\{b\}\{a,c\}$};
    \node (3) at (4, -1) [label=0:{[2,1]}] {$\{c\}\{a,b\}$};
    \node (12) at (-5, -2) [label=270:{\footnotesize $[2,1,1],\begin{matrix}010\\101\\010\end{matrix}$}] {$\{a\}\{b\}\{c\}$};
    \node (13) at (-3, -2) [label=270:{\footnotesize $[2,1,1],\begin{matrix}001\\001\\110\end{matrix}$}] {$\{a\}\{c\}\{b\}$};
    \node (21) at (-1, -2) [rectangle, draw, thick,  label=270:{\footnotesize $[2,2,2],\begin{matrix}011\\100\\100\end{matrix}$}] {$\{b\}\{a\}\{c\}$};
    \node (23) at (1, -2) [label=270:{\footnotesize $[2,2,2],\begin{matrix}011\\100\\100\end{matrix}$}] {$\{b\}\{c\}\{a\}$};
    \node (31) at (3, -2) [label=270:{\footnotesize $[2,1,1],\begin{matrix}001\\001\\110\end{matrix}$}] {$\{c\}\{a\}\{b\}$};
    \node (32) at (5, -2) [label=270:{\footnotesize $[2,1,1],\begin{matrix}010\\101\\010\end{matrix}$}] {$\{c\}\{b\}\{a\}$};

    \node (a1) at (-5.5, -5)[circle, draw] {1};
    \node (a2) at (-5, -4.14)[circle, draw] {2};
    \node (a3) at (-4.5, -5)[circle, draw] {3};

    \node (b1) at (-3.5, -5) [circle, draw] {1};
    \node (b2) at (-3, -4.14)[circle, draw] {3};
    \node (b3) at (-2.5, -5)[circle, draw] {2};

    \node (c1) at (-1.5, -5)[circle, draw] {2};
    \node (c2) at (-1, -4.14)[circle, draw] {1};
    \node (c3) at (-0.5, -5)[circle, draw] {3};

    \node (d1) at (0.5, -5)[circle, draw] {3};
    \node (d2) at (1, -4.14)[circle, draw] {1};
    \node (d3) at (1.5, -5)[circle, draw] {2};

    \node (e1) at (2.5, -5)[circle, draw] {2};
    \node (e2) at (3, -4.14)[circle, draw] {3};
    \node (e3) at (3.5, -5)[circle, draw] {1};

    \node (f1) at (4.5, -5)[circle, draw] {3};
    \node (f2) at (5, -4.14)[circle, draw] {2};
    \node (f3) at (5.5, -5)[circle, draw] {1};
  \end{scope}
  \begin{scope}
    \path [-] (0) edge (1);
    \path [-] (0) edge (2);
    \path [-] (0) edge (3);
    \path [-] (1) edge (12);
    \path [-] (1) edge (13);
    \path [-] (2) edge (21);
    \path [-] (2) edge (23);
    \path [-] (3) edge (31);
    \path [-] (3) edge (32);

    \path [-] (a1) edge (a2);
    \path [-] (a2) edge (a3);

    \path [-] (b1) edge (b2);
    \path [-] (b2) edge (b3);

    \path [-] (c1) edge (c2);
    \path [-] (c2) edge (c3);

    \path [-] (d1) edge (d2);
    \path [-] (d2) edge (d3);

    \path [-] (e1) edge (e2);
    \path [-] (e2) edge (e3);

    \path [-] (f1) edge (f2);
    \path [-] (f2) edge (f3);
  \end{scope}
\end{tikzpicture}
\end{center}
\caption{Node invariants and the canonical leaf for the search tree
  from Figure~\ref{fig:tree}. For all nodes the coloring
  $R(G, \pi_0, \nu)$, and the invariant $\phi(G, \pi_0, \nu)$ are
  printed. Colored graphs that correspond to leaves are drawn. }
\label{fig:invariant}
\end{figure} 

\end{exa}

McKay and Piperno prove the following central result (Lemma~4 in
\cite{mckay_piperno}).

\begin{thm}
  \label{thm:abstract_correctness}
  If $R$, $T$, and $\phi$ satisfy all axioms, then for the function $C$
  determined by $R$, $T$ and $\phi$ it holds that two colored graphs
  $(G, \pi_0)$ and $(G', \pi_0')$ are isomorphic if and only if
  $C(G, \pi_0) = C(G', \pi_0')$.
\end{thm}

Note that although the abstract scheme described in this section is slightly
modified compared to the original description given in \cite{mckay_piperno}
(mainly concerning the axioms for the function $\phi$ and the definition of
the canonical form), the main
result stated in Theorem~\ref{thm:abstract_correctness} still holds. A proof of the theorem in our context is provided in the appendix of this paper.
We have also formalized the described abstract scheme and
formally proved the Theorem~\ref{thm:abstract_correctness} in Isabelle/HOL.

\subsection{Search tree pruning}

Suppose that the search tree $\mathcal{T}(G,\pi_0)$ is constructed. In
order to find its canonical leaf efficiently, the search tree is
\emph{pruned} by removing the subtrees for which it is somehow deduced
that they cannot contain the canonical leaf. There are two main types
of pruning mentioned in \cite{mckay_piperno} that are of interest
here.

The first type of pruning is based on comparing node invariants -- if
for two nodes $\nu'$ and $\nu''$ such that $|\nu'| = |\nu''|$ it holds
$\phi(G,\pi_0,\nu') > \phi(G,\pi_0,\nu'')$, then the subtree
$\mathcal{T}(G,\pi_0,\nu'')$ can be pruned (operation $P_A$ in
\cite{mckay_piperno}).

The other type of pruning is based on automorphisms -- if there are
two nodes $\nu'$ and $\nu''$ such that $|\nu'|=|\nu''|$ and
$\nu' <_{lex} \nu''$ (where $<_{lex}$ denotes the lexicographic order
relation), and there is an automorphism $\sigma \in Aut(G,\pi_0)$ such
that $\nu'^\sigma = \nu''$, then the subtree
$\mathcal{T}(G,\pi_0,\nu'')$ can be pruned (operation $P_C$ in
\cite{mckay_piperno}).\footnote{In \cite{mckay_piperno}, the operation
  $P_B$ is also considered, which removes a node whose invariant is
  different from the invariant of the node on the same level on the
  path to a predefined leaf $\nu_0$. This type of pruning is only used
  to discover the automorphism group of the graph, but not for the
  discovery of its canonical form, so it is not considered in this
  work.}

In practical implementations, the search tree is not explicitly
constructed, and pruning is done as it is traversed, as soon as
possible. When a subtree is marked for pruning, its traversal is
completely avoided, which is the main source of efficiency gains. The
algorithm keeps track of the current canonical leaf candidate, which
is updated whenever a ``better'' leaf is discovered.

\begin{exa}

  The search tree in Figure~\ref{fig:invariant} can be pruned with
  either of the two types of pruning operations. For example, nodes
  with the node invariant $[2, 1]$ can be pruned, because a node with
  the node invariant $[2, 2]$ is present in the tree and
  $[2, 1] < [2, 2]$. Some nodes may also be pruned due to the
  automorphism $(a\ c)$, as shown in Figure~\ref{fig:autprune}.

\begin{figure}[h!]
\begin{center}
 \begin{tikzpicture}
  \begin{scope}
    \node (0) at (0, 0)  {$\{a,b,c\}$};
    \node (1) at (-4, -1) [label=90:$\nu_1$] {$\{a\}\{b,c\}$};
    \node (2) at (0, -1)  {$\{b\}\{a,c\}$};
    \node (3) at (4, -1) [] {$P_C(\nu_1, (a\ c))$};
    \node (12) at (-5, -2)  {$\{a\}\{b\}\{c\}$};
    \node (13) at (-3, -2)  {$\{a\}\{c\}\{b\}$};
    \node (21) at (-1, -2) [label=90:$\nu_2$] {$\{b\}\{a\}\{c\}$};
    \node (23) at (1, -2) [] {$\ \ \ \ \ \ \ \ P_C(\nu_2, (a\ c))$};

    \node (a1) at (-5.5, -3.86)[draw, circle] {1};
    \node (a2) at (-5, -3)[draw, circle] {2};
    \node (a3) at (-4.5, -3.86)[draw, circle] {3};

    \node (b1) at (-3.5, -3.86)[draw, circle] {1};
    \node (b2) at (-3, -3)[draw, circle] {3};
    \node (b3) at (-2.5, -3.86)[draw, circle] {2};

    \node (c1) at (-1.5, -3.86)[draw, circle] {2};
    \node (c2) at (-1, -3)[draw, circle] {1};
    \node (c3) at (-0.5, -3.86)[draw, circle] {3};
  \end{scope}
  \begin{scope}
    \path [-] (0) edge (1);
    \path [-] (0) edge (2);
    \path [-] (0) edge (3);
    \path [-] (1) edge (12);
    \path [-] (1) edge (13);
    \path [-] (2) edge (21);
    \path [-] (2) edge (23);

    \path [-] (a1) edge (a2);
    \path [-] (a2) edge (a3);

    \path [-] (b1) edge (b2);
    \path [-] (b2) edge (b3);

    \path [-] (c1) edge (c2);
    \path [-] (c2) edge (c3);
  \end{scope}
\end{tikzpicture}
\end{center}
\caption{The search tree from Figure~\ref{fig:invariant} pruned using
the automorphism $(a\ c)$}
\label{fig:autprune}
\end{figure} 

\end{exa}

We have formalized pruning operations in Isabelle/HOL and proved their
correctness.

\section{Proving canonical labelling and non-isomorphism}

To verify that two graphs are indeed non-isomorphic, we can either
unconditionally trust the algorithm that computed the canonical forms
(which turned out to be different), or extend the algorithm to produce
\emph{a certificate (a proof)} for each canonical form it computes,
which can be independently verified by an external tool. A certificate
is a sequence of easily verifiable steps that proves that the graph
computed by the algorithm is equal to the canonical form $C(G, \pi_0)$
as defined by the McKay and Piperno's scheme. The proof system
presented in this paper assumes that the scheme is instantiated with a
concrete refinement function, target cell selector, and node invariant
function, denoted respectively by $\bar{R}$, $\bar{T}$, and
$\bar{\phi}$.  These functions are specified in the following text,
and are similar to the functions used in real implementations.

This approach assumes the correctness of the abstract McKay and
Piperno's scheme. This meta-theory has been well-studied in the
literature \cite{mckay_practical, mckay_piperno}, and we have
formalized it in Isabelle/HOL. It is also assumed that our concrete
functions $\bar{R}$, $\bar{T}$ and $\bar{\phi}$ satisfy the
corresponding axioms (and this is explicitly proved in the following
text and in our Isabelle/HOL formalization). Therefore, the soundness
of the approach is based on Theorem~\ref{thm:abstract_correctness},
instantiated for this particular choice of $R$, $T$ and $\phi$.

On the other hand, the correctness of the implementation of the
algorithm that computes the canonical form of a given graph is neither
assumed nor verified. Instead, for the given colored graph
$(G, \pi_0)$ it generates a colored graph $(G', \pi')$ and a proof $P$
certifying that $(G', \pi')$ is the canonical form of the graph
$(G, \pi_0)$ (i.e.~$C(G,\pi_0) = (G',\pi')$).  The certificate $P$ is
verified by an independent proof-checker. Note that the implementation
of the proof-checker is much simpler than the implementation of the
algorithm for computing the canonical form, so it can be much more
easily trusted or verified within a proof assistant.

\subsection{Refinement function \texorpdfstring{$\bar{R}$}{}}

In this section we will define a specific refinement function
$\bar{R}$ and prove its correctness i.e., prove that it satisfies
axioms R1-R3.

We say that the coloring $\pi$ is \emph{equitable} if for every two
vertices $v_1$ and $v_2$ of the same color, and every color $k$ of
$\pi$, the numbers of vertices of color $k$ adjacent to $v_1$ and to
$v_2$ are equal. As in \cite{mckay_piperno}, we define the coloring
$\bar{R}(G,\pi_0,\nu)$ as the coarsest equitable coloring finer than
$\pi_0$ that individualizes the vertices from the sequence $\nu$. Such
a coloring is obtained by individualizing the vertices from $\nu$, and
then making the coloring equitable by an iterative procedure
\cite{mckay_piperno}. In each iteration we choose a cell $W$ from the
current coloring $\pi$ (called a \emph{splitting cell}), and then
partition all the cells of $\pi$ with respect to the number of
adjacent vertices from $W$. This process is repeated until an
equitable coloring is achieved. It is known \cite{mckay_piperno} that
such a coloring is unique up to the order of its cells. Thus, in order
to fully determine the coloring $\bar{R}(G,\pi_0,\nu)$, we must fix
this order in some way. The order of the cells depends on the order in
which the splitting cells are chosen and the order in which the
obtained partitions of a split cell are inserted into the resulting
coloring. In our setting, we use the following strategy:

\begin{itemize}
\item We always choose the splitting cell that is the first in the
  current coloring $\pi$ (i.e.~the cell corresponding to the smallest
  color of $\pi$) that has some effect (causes splitting of some
  cell).

\item The obtained fragments of a split cell $X$ are ordered by the
  number of adjacent vertices from the current splitting cell $W$
  (ascending). After the fragments are ordered in this way, the first
  fragment with the maximum possible size is moved to the end of the
  sequence, preserving the order of other fragments.\footnote{This
    last step is used only for the efficiency, since we want to avoid
    having the largest fragment as the splitting cell in the future.}
\end{itemize}

An efficient procedure that computes $\bar{R}(G, \pi_0, \nu)$ is given
in Algorithm~\ref{algo:refinement}. This procedure first constructs
the initial equitable coloring, and then individualizes vertices from
$\nu$ one by one, making the coloring equitable after each
step. Individualization of a vertex $v$ is done by replacing the cell
$W$ containing $v$ in the current coloring $\pi$ with two cells
$\{ v \}$ and $W \setminus \{ v \}$ in that order. In order to obtain
an equitable coloring finer than the current coloring $\pi$, the
procedure $\mathtt{make\_equitable}(G, \pi, \alpha)$ is invoked
(Algorithm~\ref{algo:calculate_coloring}), where $\alpha$ is the set
of splitting cells to use (a subset of the set of cells of $\pi$). In
case of initial equitable coloring (when $\nu$ is the empty sequence),
$\alpha$ contains all the cells of $\pi_0$. Otherwise, $\alpha$
contains only the cell $\{ v \}$, where $v$ is the vertex that is
individualized last. The procedure
$\mathtt{make\_equitable}(G, \pi, \alpha)$ implements the iterative
algorithm given in \cite{mckay_piperno}, respecting the order of cells
that is described in the previous paragraph.
    
\begin{algorithm}[!h]
\caption{$\mathtt{make\_equitable}(G, \pi, \alpha)$}
\label{algo:calculate_coloring}
\begin{algorithmic}
\REQUIRE $(G, \pi)$ is a colored graph
\REQUIRE $\alpha$ is a subset of cells from $\pi$
\ENSURE $\pi'$ is the coarsest equitable coloring of $G$ finer than $\pi$ obtained from $\pi$ by partitioning its cells
        with respect to the cells from $\alpha$
\STATE{\textbf{begin}}
 \STATE $\pi' = \pi$
 \WHILE {$\pi'$ is not discrete and $\alpha$ is not empty}
  \STATE\COMMENT{let $W$ be the first cell of $\pi'$ that belongs to $\alpha$}
  \STATE remove $W$ from $\alpha$ 
  \FOR{ each non-singleton cell $X$ of $\pi'$ }
     \STATE partition $X$ into $X_1,X_2,\ldots X_k$, according to the number of adjacent vertices from $W$
     \STATE\COMMENT{ assume that the sequence $X_1,X_2,\ldots, X_k$ is sorted by the number of adjacent vertices from $W$ (ascending)}
     \STATE\COMMENT{ let $j$ be the smallest index such that $|X_j| = max\{ |X_i|\ |\ 1 \le i \le k \}$}
     \STATE replace $X$ in $\pi'$ by $X_1,\ldots X_{j-1},X_{j+1},\ldots,X_k,X_j$ \COMMENT{in that order (\emph{in situ})}
     \FOR{ each $X_i$ in $\{ X_1, \ldots, X_{j-1}, X_{j+1}, \ldots, X_k\}$ }
       \STATE add $X_i$ to $\alpha$
     \ENDFOR
     \IF{ $X \in \alpha$ }
       \STATE replace $X$ by $X_j$ in $\alpha$
     \ENDIF
  \ENDFOR 
 \ENDWHILE 
 \STATE{ $\mathbf{return}\ \pi'$ }
\STATE{\textbf{end}}
\end{algorithmic}
\end{algorithm}

\begin{algorithm}[!h]
\caption{$\bar{R}(G, \pi_0, \nu)$}
\label{algo:refinement}
\begin{algorithmic}
\REQUIRE $(G, \pi_0)$ is a colored graph
\REQUIRE $\nu$ is a sequence of vertices of $G$
\ENSURE $\pi = \bar{R}(G, \pi_0, \nu)$
  is the coarsest equitable coloring finer than $\pi_0$ that
  individualizes vertices from $\nu$ \STATE{\textbf{begin}}
 
 \IF{$\nu = [\ ]$}
  \STATE\COMMENT{let $\alpha$ be the list of all cells of $\pi_0$}
  \STATE $\pi = \mathtt{make\_equitable}(G, \pi_0, \alpha)$
 \ELSE 
  \STATE\COMMENT{let $\nu = [\nu', v]$}
  \STATE $\pi' = \bar{R}(G, \pi_0, \nu')$
  \STATE\COMMENT{let $W$ be the cell containing $v$ in $\pi'$}
  \STATE\COMMENT{let $\pi''$ is obtained by replacing $W$ in $\pi'$ with two cells $\{ v \}, W\setminus \{ v \}$ in that order (\emph{in situ})}
  \STATE\COMMENT{let $\alpha$ contain only the cell $\{ v \}$}
  \STATE $\pi = \mathtt{make\_equitable}(G, \pi'', \alpha))$
 \ENDIF
 \STATE{$\mathbf{return}\ \pi$}
\STATE{\textbf{end}}
\end{algorithmic}
\end{algorithm}

The next two lemmas state some properties of the procedure 
$\mathtt{make\_equitable}(G, \pi, \alpha)$, needed for proving the correctness 
of the refinement function itself.

\begin{lem} 
\label{lemma:make_equitable_1}
For each $\sigma \in S_n$, 
$\mathtt{make\_equitable}(G^\sigma, \pi^\sigma, \alpha^\sigma) =
\mathtt{make\_equitable}(G,\pi,\alpha)^\sigma$.
\end{lem}

\begin{lem}
\label{lemma:make_equitable_2}

If $\pi' = \mathtt{make\_equitable}(G,\pi,\alpha)$, then
$\pi' \preceq \pi$.
\end{lem}

Using the previous two lemmas, we can prove the following lemma which
states that the refinement function $\bar{R}$ satisfies the properties
required by the McKay and Piperno's scheme.

\begin{lem}
\label{lemma:R_correctness}
If $\pi = \bar{R}(G,\pi_0,\nu)$, then:
  \begin{enumerate}
   \item $\pi \preceq \pi_0$
   \item $\pi$ individualizes vertices from $\nu$
   \item for each $\sigma \in S_n$,
     $\bar{R}(G^\sigma,\pi_0^\sigma,\nu^\sigma) =
     \bar{R}(G,\pi_0,\nu)^\sigma$
  \end{enumerate}
\end{lem}

The cells that are used for splitting are determined by the set
$\alpha$, so the algorithm does not need to perform any significant
additional work to find the next splitting cell. The next lemma shows
that the cell $W$ chosen as a splitting cell in an arbitrary iteration
of the \texttt{while} loop is indeed the first cell of $\pi'$ for
which the partitioning of $\pi'$ with respect to $W$ would possibly
have some effect on $\pi'$.

\begin{lem}
 \label{lemma:make_equitable}
 Let $W$ be the cell of $\pi'$ that is chosen as a splitting cell in
 some arbitrary iteration of the \texttt{while} loop of the procedure
 $\mathtt{make\_equitable}(G, \pi, \alpha)$. Let $W'$ be any cell of
 $\pi'$ that precedes $W$ in $\pi'$ (that is, it corresponds to a
 smaller color). Then $W'$ does not cause any splitting, i.e.,~for any
 two vertices of a same color in $\pi'$ the numbers of their adjacent
 vertices from $W'$ are equal.
\end{lem}

A naive implementation of the \texttt{make\_equitable} procedure does
not use the set $\alpha$ and tries to find the splitting cell by
probing cells. It iterates through cells of $\pi'$ (in order) trying
to use them for splitting either until splitting on the current cell
has some effect (some cell is split), or until all cells are processed
with no effect, proving that the current coloring is equitable. We use
that simpler procedure within our proof system and proof checker
(since it is simpler than the efficient one) and we have formally
proved its correctness in Isabelle/HOL.
  
\subsection{Target cell selector \texorpdfstring{$\bar{T}$}{}}

We assume that the target cell $\bar{T}(G,\pi_0,\nu)$ that corresponds
to the node $\nu$ of the search tree is the first non-singleton cell
of the coloring $\bar{R}(G,\pi_0,\nu)$ if such a cell exists, or the
empty set, otherwise (this only happens in leaves). It trivially
satisfies the axioms T1 and T2, while the following lemma states that
the target cell chosen this way is label invariant, and therefore
satisfies the axiom T3.

\begin{lem}
  \label{lemma:T_label_invariant}
  For each $\sigma \in S_n$ it holds
  $\bar{T}(G^\sigma, \pi_0^\sigma, \nu^\sigma) = \bar{T}(G,\pi_0,\nu)^\sigma$.
\end{lem}

We have formalized $\bar{T}$ and its properties in Isabelle/HOL.

\subsection{Node invariant \texorpdfstring{$\bar{\phi}$}{}}

Node invariants $\bar{\phi}$ are defined using a suitable function
on colored graphs (denoted by $\operatorname{f}_{hash}(G,\pi)$)
which is label invariant, i.e.,
$\operatorname{f}_{hash}(G^\sigma,\pi^\sigma) =
\operatorname{f}_{hash}(G,\pi)$ for each permutation $\sigma \in
S_n$. Our approach to construct such function is based on
\emph{quotient graphs} \cite{mckay_piperno}. The quotient graph that
corresponds to a colored graph $(G,\pi)$ is the graph whose vertices
are the cells of $\pi$ labeled by the cell number and size, and the
edge that connects any two cells $W_1$ and $W_2$ is labeled by the
number of edges between the vertices of $W_1$ and $W_2$ in $G$. It is
easy to prove that quotient graphs are label invariant, so any hash
function that depends only on the quotient graph will also be label
invariant. In our setting, the node invariant
$\bar{\phi}(G,\pi_0,\nu)$ that corresponds to the node
$\nu = [v_1,v_2,\ldots,v_k]$ is a vector of numbers
$[h_1,h_2,\ldots,h_k]$, such that $h_i = \operatorname{f}_{hash}(G,R(G,\pi_0,[v_1,v_2,\ldots,v_i]))$ for each $i$ ($1 \le i \le k$). The function $\operatorname{f}_{hash}$ first constructs the quotient graph,
and then applies some fixed hash function to it. The node invariants
$\bar{\phi}$ are ordered lexicographically.
       
The following lemma proves that the axioms $\phi$1-$\phi$3 are
satisfied for such defined node invariant function.

\begin{lem}
  \label{lemma:phi_correct}
  The following properties of the function $\bar{\phi}$ hold:
\begin{enumerate}
\item If $|\nu'| = |\nu''|$ and
  $\bar{\phi}(G,\pi_0,\nu') <
  \bar{\phi}(G,\pi_0,\nu'')$, then
  $\bar{\phi}(G,\pi_0,[\nu', w_1]) <
  \bar{\phi}(G,\pi_0,[\nu'',w_2])$, for each
  $w_1 \in T(G,\pi_0,\nu')$,
  $w_2 \in T(G,\pi_0,\nu'')$
\item For each $\nu$ and nonempty $\nu'$ such that
  $[\nu, \nu']$ is a tree node, it holds that
  $\bar{\phi}(G, \pi_0, \nu) < \bar{\phi}(G, \pi_0, [\nu, \nu'])$.
\item
  $\bar{\phi}(G^\sigma, \pi_0^\sigma, \nu^\sigma) =
  \bar{\phi}(G,\pi_0,\nu)$ for each node $\nu$ and for each permutation
  $\sigma \in S_n$
\end{enumerate}
\end{lem}

We have formalized the notion of quotient graphs, defined the function
$\bar{\phi}$ and formally proved its properties in Isabelle/HOL.

\subsection{The proof system}
\label{subsec:the_proof_system}

\emph{A proof} is a sequence of \emph{rule applications}, each
deriving one or more \emph{facts} from the \emph{premises}, which are
the facts already derived by the preceding rules in the sequence.
The proof can be built during the search tree traversal by exporting
rule applications determined based on the operations performed during
the algorithm execution. However, it is possible to optimize this
approach and emit proofs after the search tree has been completely
processed -- such proofs can be much shorter, avoiding derivation of
many facts that are not relevant for the final result. This will be
discussed in more details in Section~\ref{subsec:impl}.

The derived facts describe search tree nodes and their properties. We
consider the following types of facts (and their associated
semantics):

\begin{itemize}
\item $\nu \in \mathcal{T}(G,\pi_0)$: the sequence $\nu$ is a node of the tree $\mathcal{T}(G,\pi_0)$
\item $\bar{R}(G,\pi_0,\nu) = \pi$: the coloring
  $\bar{R}(G, \pi_0, \nu)$ that corresponds to the node $\nu$ is equal
  to $\pi$
\item $\pi \xrightarrow{eq} \bar{R}(G,\pi_0,\nu)$: the coloring
  $\bar{R}(G, \pi_0, \nu)$ that corresponds to the node $\nu$ is equal
  to the coloring obtained by refining the coloring $\pi$, i.e., by
  invoking $\texttt{make\_equitable}(G,\pi,\alpha)$, where $\alpha$ is
  the list of all cells of $\pi$
\item $\bar{T}(G,\pi_0,\nu) = W$: the target cell
  $\bar{T}(G,\pi_0,\nu)$ that corresponds to the node $\nu$ is $W$,
  which is a non-empty set of vertices from $G$
\item $\Omega \vartriangleleft \operatorname{orbits}(G, \pi_0, \nu)$:
  the set of graph vertices $\Omega$ is a subset of some orbit with respect to
  $Aut(G,\pi)$, where $\pi = \bar{R}(G,\pi_0,\nu)$.
\item
  $\bar{\phi}(G,\pi_0, \nu') =
  \bar{\phi}(G,\pi_0,\nu'')$: the node invariants that
  correspond to the nodes $\nu'$ and $\nu''$ (where $|\nu'|=|\nu''|$) are equal
\item $\operatorname{pruned}(G,\pi_0, \nu)$: the node
  $\nu$ is not an ancestor of the canonical leaf
\item $\operatorname{on\_path}(G,\pi_0,\nu)$: the node
  $\nu$ is an ancestor of the canonical leaf
\item $C(G,\pi_0) = (G',\pi')$: the colored graph $(G',\pi')$
  is the canonical form of the colored graph $(G,\pi_0)$
\end{itemize}

The goal of the proving process is to derive a fact of the form
$C(G,\pi_0) = (G',\pi')$ (i.e.~the final rule in the proof sequence
should derive such a fact).

While some rules may be applied unconditionally, whenever
facts in their premises are already derived, there are also rules that
require some additional conditions to be fulfilled in order to be
applied. These additional conditions should be easily checkable during
the proof verification. The following operations are used for
formulating and checking those conditions (colorings are represented
by lists of their cells):

\begin{itemize}
\item $\operatorname{cell}_j(\pi)$ -- the $j$-th cell of $\pi$
\item $|\operatorname{cell}_j(\pi)|$ -- size of the $j$-th cell of $\pi$
\item $\operatorname{discrete}(\pi)$ -- true if and only if the
  coloring $\pi$ is discrete
\item $\operatorname{individualize}(\pi,v)$ -- the coloring
  obtained from $\pi$ by individualizing the graph vertex $v$, i.e.~by
  replacing the cell $W$ of $\pi$ containing $v$ by two cells
  $\{ v \}$ and $W \setminus \{ v \}$ in that order (\emph{in situ})
\item $\operatorname{split}(G,\pi,i)$ -- the coloring obtained by
  partitioning the cells of $\pi$ with respect to the number of
  adjacent vertices from the $i$-th cell of $\pi$. Each cell $X$ of
  $\pi$ is replaced by the cells $X_1,\ldots,X_k$ obtained by its
  partitioning, ordered in the same way as in
  Algorithm~\ref{algo:calculate_coloring}
\item $\operatorname{split}(G, \pi, i) \prec \pi$ -- the coloring
  obtained by partitioning the cells of $\pi$ with respect to the
  number of adjacent vertices from the $i$-th cell of $\pi$ is
  strictly finer than $\pi$ (some cells are split)
\item $\operatorname{split}(G, \pi, i) = \pi$ -- the coloring obtained
  by partitioning the cells of $\pi$ with respect to the number of
  adjacent vertices from the $i$-th cell of $\pi$ is equal to $\pi$
  (no cells are split)
\item $\operatorname{f}_{hash}(G,\pi)$ -- the hash value that
  corresponds to the colored graph $(G,\pi)$ (i.e., to its quotient
  graph)
\item $v^\pi$ -- action of the permutation $\pi$ on the graph vertex $v$
\item $G^\pi$ -- action of the permutation $\pi$ on the graph $G$
\item $\pi_0^\pi$ -- action of the permutation $\pi$ on the coloring $\pi_0$
\item $G^{\pi_1} > G^{\pi_2}$ -- true if and only if the graph $G$ permuted by the
  permutation $\pi_1$ is greater than the graph $G$ permuted by the
  permutation $\pi_2$ in the lexicographic order of their adjacency
  matrices.
\item $\nu' <_{lex} \nu''$ -- true if and only if the list of numbers
  $\nu'$ is lexicographically smaller than the list of numbers $\nu''$
\item $\sigma \in Aut(G, \pi_0)$ -- true if and only if $\sigma$ is an
  automorphism of of the colored graph $(G, \pi_0)$
\end{itemize}

Note that most of these operations are also used in the canonical
form algorithm implementation, so their implementation might be shared
between that implementation and the implementation of the proof
checker. However, if the checker is not mechanically verified, it is
better to use less efficient, but more straightforward implementation
of those operations (since they must be trusted or verified only by a
manual code inspection). They are less efficient, but the number of
their calls is usually much smaller in the checker than in the
canonical form finding algorithm itself.

The rules of our proof system will be described in the following
sections. When the rules are printed, premises will be displayed above a horizontal
line, derived facts below a line, and additional conditions will be
printed below (within \textbf{where} and \textbf{provided} clauses).

\subsubsection{Refinement rules}

The rules given in Figure~\ref{rules:refinement} are used to verify
the correctness of the colorings constructed by the refinement
function $\bar{R}$.

The rule $\mathtt{ColoringAxiom}$ formally derives that the empty
sequence is a node of the search tree (it is its root), and that the
equitable coloring assigned to this node is obtained by applying the
function $\texttt{make\_equitable}$ (Algorithm~\ref{algo:calculate_coloring}) to the initial coloring $\pi_0$. This
rule is exported once, at the very beginning of the proof
construction.

The rule $\mathtt{Individualize}$ verifies colorings obtained by
individualizing vertices (an instance of this rule can be exported
after each individualization in Algorithm~\ref{algo:refinement}).

The rule $\mathtt{SplitColoring}$ verifies colorings obtained by
partitioning with respect to a selected cell (an instance of this rule
can be exported in each iteration of the $while$ loop in
Algorithm~\ref{algo:calculate_coloring}). Note that this rule has an
additional condition that ensures that the correct splitting cell is
used. This is ensured by explicitly checking that the splitting on
any previous cell of $\pi$ has no effect on $\pi$
(this is in agreement with Lemma~\ref{lemma:make_equitable}).

The rule $\mathtt{Equitable}$ formally derives that an equitable
coloring is achieved. This is ensured by checking explicitly that
splitting on any cell of $\pi$ has no effect. This rule can be
exported at the end of Algorithm~\ref{algo:calculate_coloring}.

\begin{figure}[!h]
{\footnotesize
$$
\begin{array}{l}
 \mathtt{ColoringAxiom}: \\
  \begin{array}{c}
       \\
  \hline
  \pi_0 \xrightarrow{eq} \bar{R}(G, \pi_0, [\ ]) \\
  \ [\ ] \in \mathcal{T}(G,\pi_0) \\
   \end{array} \\
   \\
 \mathtt{Individualize}(\nu,v,\pi): \\
  \begin{array}{c}
  \bar{R}(G, \pi_0, \nu) = \pi \\
  \, [\nu, v] \in \mathcal{T}(G,\pi_0) \\
  \hline
  \operatorname{individualize}(\pi, v) \xrightarrow{eq} \bar{R}(G, \pi_0, [\nu, v])\\
   \end{array} \\
   \\
 \mathtt{SplitColoring}(\nu,\pi):\\
  \begin{array}{c}
  \pi \xrightarrow{eq} \bar{R}(G, \pi_0, \nu) \\
  \hline
  \operatorname{split}(G, \pi, i) \xrightarrow{eq} \bar{R}(G, \pi_0, \nu)
   \end{array} \\
    \left\{\begin{array}{l}
    \textbf{where: } i = \min\{ j\ |\ \operatorname{split}(G, \pi, j) \prec \pi \} \\
    \textbf{provided: }\text{such $i$ exists} \\
  \end{array}\right.  \\ 
  \\
\mathtt{Equitable}(\nu,\pi): \\
  \begin{array}{c}
  \pi \xrightarrow{eq} \bar{R}(G, \pi_0, \nu) \\
  \hline
  \bar{R}(G, \pi_0, \nu) = \pi
   \end{array}  \\
  \left\{\begin{array}{l}
    \textbf{provided: }  \forall i.\ \operatorname{split}(G, \pi, i) = \pi \\
  \end{array}\right. \\
\end{array}
$$
}
\caption{Rules for refinement}
\label{rules:refinement}
\end{figure}

Note that $\mathtt{SplitColoring}$ rule applications explicitly encode
intermediate colorings obtained at each iteration of the \emph{while} loop in
the $\texttt{make\_equitable}$ procedure. The proof format could be
changed to make the proof objects much smaller by omitting
$\mathtt{SplitColoring}$ rule applications. In that case, the checker
would have to run the full $\texttt{make\_equitable}$ procedure (or at
least its naive implementation) when checking the modified
$\mathtt{Equitable}$ rule (shown in Figure~\ref{rules:equitable}).

\begin{figure}[!h]
{\footnotesize
$$
\begin{array}{l}
\mathtt{Equitable}(\nu,\pi,\pi'): \\
  \begin{array}{c}
  \pi \xrightarrow{eq} \bar{R}(G, \pi_0, \nu) \\
  \hline
  \bar{R}(G, \pi_0, \nu) = \pi'
   \end{array}  \\
  \left\{\begin{array}{l}
           \textbf{provided: }  \mathtt{make\_equitable}(G, \pi_0, \alpha) = \pi',\\
           \textrm{where}\ \alpha\ \textrm{contains\ all\ cells\ of}\ \pi\\
  \end{array}\right. \\
\end{array}
$$
}
\caption{Modified $\mathtt{Equitable}$ rule}
\label{rules:equitable}
\end{figure}

\subsubsection{Rule for the target cell selection}
 
The selection of correct target cells is verified by the rule
$\mathtt{TargetCell}$, which is shown in Figure~\ref{rules:target}.
This is ensured by explicitly checking that all previous cells are
singleton, while the selected target cell has more than one element.
An instance of this rule can be exported by the algorithm whenever the
target cell selector is applied. Note that this rule also derives the
facts which state that the sequences of the form $[\nu, v]$, where $v$
belongs to the target cell corresponding to the node $\nu$, are also
nodes of the search tree. Such facts are important premises for the
application of several other rules.

 \begin{figure}[!h]
{\footnotesize
$$
\begin{array}{l}
 \mathtt{TargetCell}(\nu,\pi): \\
  \begin{array}{c}
  \bar{R}(G, \pi_0, \nu) = \pi \\
  \hline
  \bar{T}(G, \pi_0, \nu) = \operatorname{cell}_i(\pi) \\
  \left[\nu, v\right] \in \mathcal{T}(G,\pi_0),\ \mathrm{for\ all}\ v \in \operatorname{cell}_i(\pi)\\
  \end{array} \\
   \left\{\begin{array}{l}
   \textbf{where: } i = \min \{ j\ |\ |\operatorname{cell}_j(\pi)| > 1 \} \\
   \textbf{provided: }\text{such $i$ exists} \\
   \end{array}\right.
   \\
\end{array}
$$}
\caption{Rule for target cell}
\label{rules:target}
\end{figure}
 
\subsubsection{Rules for the node invariant equality detection}
 
The rules listed in Figure~\ref{rules:invariant} are used for
verifying that some nodes at the same level of the search tree have
equal node invariants (recall that these are sequences of hash
values). Those facts are important for verifying the lexicographic
order over the node invariants, and finding the maximal leaf which
yields the canonical form.

The rule $\mathtt{InvariantAxiom}$ derives facts using the reflexivity
of the equality. It can be exported once for each node of the tree.

The rule $\mathtt{InvariantsEqualSym}$ derives facts using the
symmetry of the equality. It can be exported for any two nodes at the
same level of the tree for which the fact of invariant equality has
already been derived.

The rule $\mathtt{InvariantsEqual}$ verifies that the node invariants
(which are sequences of hash values) of two nodes at the same level
are equal -- their parents should have equal node invariants and the
hash values assigned to the graph $G$ colored by corresponding
colorings of the two nodes should be equal (this is explicitly checked
in the checker by computing $\operatorname{f}_{hash}$ values). This
rule may be exported whenever two nodes at the same level in the tree
have equal node invariants.

\begin{figure}[!h]
{\footnotesize $$
\begin{array}{l}
 \mathtt{InvariantAxiom}(\nu): \\
  \begin{array}{c}
   \nu \in \mathcal{T}(G,\pi_0) \\
  \hline
   \bar{\phi}(G,\pi_0, \nu) = \bar{\phi}(G,\pi_0, \nu)
   \end{array}
  \\
  \\
   \mathtt{InvariantsEqualSym}(\nu',\nu''): \\
   \begin{array}{c}
   \bar{\phi}(G,\pi_0,\nu') = \bar{\phi}(G,\pi_0,\nu'') \\
   \hline 
    \bar{\phi}(G,\pi_0,\nu'') = \bar{\phi}(G,\pi_0,\nu') \\
   \end{array}
  \\
  \\
 \mathtt{InvariantsEqual}(\nu',v',\pi_1,\nu'',v'',\pi_2): \\
  \begin{array}{c}
  \bar{\phi}(G,\pi_0, \nu') = \bar{\phi}(G,\pi_0,\nu''), \\
  \bar{R}(G, \pi_0, [\nu',v']) = \pi_1 \\
  \bar{R}(G, \pi_0, [\nu'', v'']) = \pi_2\\
  \hline
   \bar{\phi}(G,\pi_0, [\nu',v']) = \bar{\phi}(G,\pi_0, [\nu'', v''])
   \end{array} \\ 
   \left\{\begin{array}{l}
     \textbf{provided: }\operatorname{f}_{hash}(G,\pi_1) = \operatorname{f}_{hash}(G,\pi_2) \\
   \end{array}
   \right.\\
   \\ 
\end{array}
$$}
\caption{Node invariant equality rules}
\label{rules:invariant}
\end{figure} 

\subsubsection{Rules for orbits calculation}

Orbit calculation is verified by rules given in Figure~\ref{rules:orbits}.

The rule $\mathtt{OrbitsAxiom}$ states that any singleton set of
vertices is a subset of some of the orbits with respect to
$Aut(G,\pi)$ (where $\pi = \bar{R}(G,\pi_0,\nu)$).  It may be exported
once for each graph vertex and for each tree node.

The rule $\mathtt{MergeOrbits}$ is used for merging the orbits
$\Omega_1$ and $\Omega_2$ whenever two vertices $w_1 \in \Omega_1$ and
$w_2\in \Omega_2$ correspond to each other with respect to some
automorphism $\sigma \in Aut(G,\pi)$. It can be exported whenever two
orbits are merged at some tree node, after a new automorphism $\sigma$
had been discovered.

\begin{figure}[!h]
{\footnotesize $$
\begin{array}{l}
 \mathtt{OrbitsAxiom}(v,\nu): \\
  \begin{array}{c}
   \nu \in \mathcal{T}(G,\pi_0) \\
  \hline
   \{ v \} \vartriangleleft \operatorname{orbits}(G,\pi_0,\nu) \\
   \end{array} \\
   \left\{\begin{array}{l}
            \textbf{provided: } v \in G \\
          \end{array}\right.\\
   \\
  \mathtt{MergeOrbits}(\Omega_1,\Omega_2,\nu,\sigma,w_1,w_2): \\
  \begin{array}{c}
    \Omega_1 \vartriangleleft \operatorname{orbits}(G,\pi_0, \nu),\\
    \Omega_2 \vartriangleleft \operatorname{orbits}(G,\pi_0, \nu),\\
  \hline
   \Omega_1 \cup \Omega_2 \vartriangleleft \operatorname{orbits}(G,\pi_0,\nu) \\
   \end{array} \\
   \left\{\begin{array}{ll}
            \textbf{provided: } &  \exists \sigma \in Aut(G,\pi_0).\ \nu^\sigma = \nu\ \wedge\\
             & \exists w_1\in \Omega_1.\ \exists w_2\in\Omega_2.\ w_1^\sigma = w_2\\
   \end{array}\right. \\
\end{array}
$$}
\caption{Orbits calculation rules}
\label{rules:orbits}
\end{figure} 

\subsubsection{Rules for pruning}

The rules given in Figure~\ref{rules:pruning} are used for
formalization of pruning. Note that the pruning within the proof
(i.e.~application of the pruning rules) has quite different purpose,
compared to the pruning during the search. Namely, the sole purpose of
pruning during the search (as described in \cite{mckay_piperno}) is to
make the tree traversal more efficient, by avoiding the traversal of
unpromising branches. In that sense, there is no use of pruning a
subtree that has been already traversed, or a leaf that has been
already visited. In other words, we can think of pruning during the
search as a synonym for \emph{not traversing}.  On the other hand,
pruning within the proof (which will be referred to as \emph{formal
  pruning}) has the purpose of \emph{proving} that some node is not an
ancestor of the canonical leaf.  This must be done for all such nodes
(whether they are traversed during the search or not), in order to be
able to prove that the remaining nodes \emph{are} the ancestors of the
canonical leaf, i.e.~that they belong to the path from the root of the
tree to the canonical leaf. This enables deriving the canonical form,
which is done by the rules described in the next section.

As a consequence, applications of the pruning rules within the proof
may or may not correspond to effective pruning operations during the
search. Some formal pruning derivations will be performed
\emph{retroactively}, i.e.~after the corresponding subtree has been
already traversed. Also, it will be necessary to prune (non-canonical)
leaves when they are visited.

 \begin{figure}[!h]
{\footnotesize $$
\begin{array}{l}
 \mathtt{PruneInvariant}(\nu',v',\pi_1,\nu'',v'',\pi_2): \\
  \begin{array}{c}
  \bar{\phi}(G,\pi_0, \nu') = \bar{\phi}(G,\pi_0,\nu''), \\
  \bar{R}(G, \pi_0, [\nu',v']) = \pi_1 \\
  \bar{R}(G, \pi_0, [\nu'', v'']) = \pi_2\\
  \hline
  \operatorname{pruned}(G,\pi_0,[\nu'',v''])
   \end{array} \\
   \left\{\begin{array}{l}
           \textbf{provided: } \operatorname{f}_{hash}(G,\pi_1) > \operatorname{f}_{hash}(G,\pi_2) \\
          \end{array}
    \right. 
   \\
   \\
 \mathtt{PruneLeaf}(\nu',\pi_1,\nu'',\pi_2): \\
  \begin{array}{c}
  \bar{R}(G,\pi_0,\nu') = \pi_1 \\
  \bar{R}(G,\pi_0,\nu'') = \pi_2,\\
  \bar{\phi}(G,\pi_0, \nu') = \bar{\phi}(G,\pi_0,\nu'') \\
  \hline
  \operatorname{pruned}(G,\pi_0, \nu'')
   \end{array} \\
   \left\{\begin{array}{l}
            \textbf{provided: }\operatorname{discrete}(\pi_2) \wedge (\neg \operatorname{discrete}(\pi_1) \vee G^{\pi_1} > G^{\pi_2}) \\
          \end{array}
\right.\\
\\
  \mathtt{PruneAutomorphism}(\nu',\nu'',\sigma): \\
  \begin{array}{c}
  \nu' \in \mathcal{T}(G,\pi_0)\\
  \nu'' \in \mathcal{T}(G,\pi_0)\\
  \hline
  \operatorname{pruned}(G,\pi_0,\nu'')
   \end{array} \\  
   \left\{\begin{array}{ll}
            \textbf{provided: }& \nu' <_{lex} \nu''\ \wedge\\
                               & \sigma \in Aut(G,\pi_0) \wedge \ \nu'^\sigma = \nu'' \\
          \end{array}
\right.\\
\\   
     \mathtt{PruneOrbits}(\Omega,\nu,w_1,w_2): \\
  \begin{array}{c}
  \, [\nu, w_1] \in \mathcal{T}(G,\pi_0) \\
  \, [\nu, w_2] \in \mathcal{T}(G,\pi_0) \\
  \Omega \vartriangleleft \operatorname{orbits}(G, \pi_0, \nu) \\
  \hline
  \operatorname{pruned}(G,\pi_0,[\nu, w_2])
   \end{array} \\
   \left\{\begin{array}{l}
           \textbf{provided: } w_2\in \Omega \wedge \exists w_1 \in \Omega.\ w_1 < w_2 \\
          \end{array}
  \right.\\
  \\
    \mathtt{PruneParent}(\nu,W): \\
  \begin{array}{c}
  \bar{T}(G,\pi_0,\nu) = W,\\
  \forall w \in W.\ \operatorname{pruned}(G,\pi_0,[\nu,w])\\
  \hline
  \operatorname{pruned}(G,\pi_0,\nu) \\
  \end{array}
\end{array}
$$}
\caption{Pruning rules}
\label{rules:pruning}
\end{figure} 

The rule $\mathtt{PruneInvariant}$ verifies the invariant based
pruning (operation $P_A$ in \cite{mckay_piperno}). It justifies
pruning of a node $\nu''$ (and its subtree) when another node $\nu'$
with the greater value of the node invariant is found on the same
level of the search tree. Note that the pruned node can also be a
leaf. Since node invariants are compared lexicographically it suffices
to show that parent nodes of $\nu'$ and $\nu''$ have equal node
invariants, and that $\nu''$ has a smaller hash value than $\nu'$
(this is explicitly checked in the checker by computing and comparing
hash values $\operatorname{f}_{hash}$). This rule may be exported when
such a pruning operation is done during the search. However, this rule
may also perform the retroactive pruning, in case when the node
$\nu''$ being pruned belongs to the path from the root of the tree to
the current canonical node candidate (which has been already
traversed), and the node $\nu'$ is a node which is visited later, but
it has a greater node invariant (that is the moment when the algorithm
performing the search updates the current canonical node candidate).

The rule $\mathtt{PruneLeaf}$ does not correspond to any type of
pruning described in \cite{mckay_piperno} as such (since no leaves are
pruned during the search). It justifies pruning of a leaf, based on
some special cases not covered by $\mathtt{PruneInvariant}$ rule.
\begin{itemize}
\item The first case is when a pruned leaf $\nu''$ has equal node
  invariant as some non-leaf node $\nu'$ such that
  $|\nu'|=|\nu''|$. In that case, it is obvious that all the leaves
  belonging to the subtree $\mathcal{T}(G,\pi_0,\nu')$ have greater
  node invariants than the leaf $\nu''$, since node invariants are
  compared lexicographically.
\item The second case is when the node $\nu'$ is also a leaf with an
  equal node invariant as the pruned leaf $\nu''$, but its
  corresponding graph is greater than the graph that corresponds to
  $\nu''$ (which is explicitly checked by the checker).
\end{itemize}
In both cases, we can prune the leaf $\nu''$ since it is not the
canonical leaf (recall that the canonical leaf has the maximal value of
the node invariant and has the maximal graph among such leaves).

The rules $\mathtt{PruneAutomorphism}$ and $\mathtt{PruneOrbits}$
verify the automorphism based pruning (operation $P_C$ in
\cite{mckay_piperno}). In $\mathtt{PruneAutomorphism}$ rule, the
checker needs to verify that the given permutation $\sigma$ is an
automorphism of $(G, \pi_0)$, that it maps the list $\nu'$ to $\nu''$,
and that $\nu'$ is lexicographically smaller than $\nu''$. The
connection of the rule $\mathtt{PruneOrbits}$ to the automorphisms is
more subtle. Namely, the fact
$\Omega \vartriangleleft \operatorname{orbits}(G,\pi_0,\nu)$ implies
that there is an automorphism $\sigma \in Aut(G,\pi_0)$ such that
$[\nu,w_1]^\sigma = [\nu, w_2]$ and $[\nu, w_1]$ is lexicographically
smaller than $[\nu, w_2]$ (since $w_1 < w_2$). Therefore, this kind of
pruning is also an instance of $P_C$ operation described in
\cite{mckay_piperno}.

Finally, the rule $\mathtt{PruneParent}$ derives the fact that a node
cannot be an ancestor of the canonical leaf because none of its
children is an ancestor of the canonical leaf.  This rule performs
retroactive pruning and plays an essential role in pruning already
traversed branches which are shown not the contain the canonical leaf.

\subsubsection{Rules for discovering the canonical leaf}  

The rules given in Figure~\ref{rules:canonical} are used to formalize
the traversal of the remaining path of the search tree in order to
reach the canonical leaf, after all the pruning is done. The rule
$\mathtt{PathAxiom}$ states that the root node of the tree belongs to
the path leading to the canonical leaf. It is exported once, when all
the pruning is finished. The rule $\mathtt{ExtendPath}$ is then
exported for each node on the path leading to the canonical leaf, and
it states that if the node $\nu$ is on that path, and all its children
except $[\nu,w]$ are pruned, then $[\nu,w]$ is also on that
path. Finally, the rule $\mathtt{CanonicalLeaf}$ is exported at the
very end of the proof construction, and it states that any leaf that
belongs to the path that leads to the canonical leaf must be the
canonical leaf itself (and, therefore, it corresponds to the canonical
graph).

 \begin{figure}[!h]
{\footnotesize $$
\begin{array}{l}
  \mathtt{PathAxiom:} \\
  \begin{array}{c}
   \\
  \hline
  \operatorname{on\_path}(G,\pi_0,[\ ])
   \end{array} \\
   \\
 \mathtt{ExtendPath}(\nu,W,w): \\
  \begin{array}{c}
   \operatorname{on\_path}(G,\pi_0,\nu) \\
   \bar{T}(G,\pi_0,\nu) = W,\\
   \forall w' \in W \setminus \{ w \}.\ \operatorname{pruned}(G,\pi_0,[\nu,w']) \\
  \hline
  \operatorname{on\_path}(G,\pi_0,[\nu,w])
   \end{array} \\
   \left\{\begin{array}{l}
           \textbf{provided: } w \in W
          \end{array}\right.\\

   \\
 \mathtt{CanonicalLeaf}(\nu,\pi): \\
  \begin{array}{c}
   \operatorname{on\_path}(G,\pi_0,\nu) \\
   \bar{R}(G,\pi_0,\nu) = \pi \\
  \hline
  C(G,\pi_0) = (G^\pi,\pi_0^\pi)
   \end{array} \\
   \left\{\begin{array}{l}
     \textbf{provided: }\operatorname{discrete}(\pi)\\
   \end{array}\right.\\
\end{array}
$$}
\caption{Canonical leaf rules}
\label{rules:canonical}
\end{figure}

We have formalized all the facts and their semantics, and all listed
rules in Isabelle/HOL.
 
\subsection{The correctness of the proof system}
 
In this section, we prove the correctness of the presented proof
system. More precisely, we want to prove the \emph{soundness}
(i.e.~that we can only derive correct canonical forms), and the
\emph{completeness} (i.e.~that we can derive the canonical form of any
colored graph). Detailed proofs are given in the Appendix, and the
soundness proofs are also formalized in Isabelle/HOL. Note that
the verified proofs of canonical forms only confirm that the derived
canonical forms are correct, without any conclusions about the
isomorphism of the given graphs. As already noted, we still rely on
the correctness of McKay and Piperno's abstract algorithm itself, that
is, that the canonical forms returned by the algorithm are equal if
and only if the two graphs are isomorphic (we have also formalized
this result in Isabelle/HOL).

\subsubsection{Soundness}

For soundness, we need to prove that if our proof derives the fact
$C(G,\pi_0) = (G',\pi')$, then $(G',\pi')$ is indeed the canonical
form of $(G,\pi_0)$ that should be returned by the instance of McKay's
algorithm described in the previous sections. We prove the soundness
of the proof system by proving the soundness of all its rules. We say
that a fact is \emph{valid} for the graph $(G,\pi_0)$ if its
associated semantics (as defined in
Section~\ref{subsec:the_proof_system}) holds in $(G,\pi_0)$. The
soundness of the rules means that they derive valid facts from valid
facts, which is the subject of the following lemma.

\begin{lem}
 \label{lemma:sound_rules}
 For each of the rules in our proof system, if all its premises are valid facts for the
 graph $(G,\pi_0)$, and all additional conditions required by the rule are fulfilled, then
 the facts that are derived by the rule are also valid facts for the graph $(G,\pi_0)$.
\end{lem}

Soundness of the rules implies the validity of all derived
facts, which is stated by the following lemma.

\begin{lem}
 \label{lemma:valid_facts}
For any proof that corresponds to the graph $(G,\pi_0)$, all the
facts that it derives are valid facts for $(G,\pi_0)$.
\end{lem}

The immediate consequence of the previous lemma is the following theorem.

\begin{thm}
  \label{thm:soundness}
  Let $(G,\pi_0)$ be a colored graph. Assume a proof that corresponds
  to $(G, \pi_0)$ such that it derives the fact
  $C(G,\pi_0)=(G',\pi')$. Then $(G', \pi')$ is the canonical
  form of the graph $(G,\pi_0)$.\qed
\end{thm}

\subsubsection{Completeness}

The completeness of the proof system means that for any colored graph
there is a proof that derives its canonical form. First, we prove the
following lemmas which claim the completeness of the proof system with
respect to particular types of facts.

\begin{lem}
\label{lemma:comp_R_T_node}
Let $\nu$ be a node of the search tree $\mathcal{T}(G,\pi_0)$ and let $\pi$ be the coloring assigned to $\nu$. Then the
facts $\nu \in \mathcal{T}(G,\pi_0)$ and $\bar{R}(G,\pi_0,\nu) = \pi$ can be
derived in our proof system. Furthermore, if $\pi$ is not discrete, and $W$
is the first non-singleton cell of $\pi$, then the fact $\bar{T}(G,\pi_0,\nu)=W$
can also be derived.
\end{lem}

\begin{lem}
\label{lemma:comp_phi}
Let $\nu_1$ and $\nu_2$ be two nodes such that $|\nu_1|=|\nu_2|$, and
that have equal node invariants. Then there is a proof that derives
the fact $\bar{\phi}(G,\pi_0,\nu_1)=\bar{\phi}(G,\pi_0,\nu_2)$.
\end{lem}

\begin{lem}
\label{lemma:comp_leaf}
If $\nu$ is a non-canonical leaf of
$\mathcal{T}(G,\pi_0)$, then there is a proof that derives the fact
$\operatorname{pruned}(G,\pi_0,[\nu]_i)$ for some $i \le |\nu|$.
\end{lem}

\begin{lem}
\label{lemma:comp_tree}
If a subtree $\mathcal{T}(G,\pi_0,\nu)$ does not contain
the canonical leaf, then there is a proof that derives the fact
$\operatorname{pruned}(G,\pi_0,[\nu]_i)$ for some $i \le |\nu|$.
\end{lem}

\begin{lem}
\label{lemma:comp_path}
If a node $\nu$ is an ancestor of the canonical leaf of the tree
$\mathcal{T}(G,\pi_0)$, then there is a proof that derives the fact
$\operatorname{on\_path}(G,\pi_0,\nu)$.
\end{lem}

\begin{lem}
\label{lemma:comp_canon}
If $\pi^*$ is the coloring that corresponds to the canonical leaf
$\nu^*$ of the tree $\mathcal{T}(G,\pi_0)$, then there is a proof that
derives the fact $C(G,\pi_0) = (G^{\pi^*},\pi_0^{\pi^*})$.
\end{lem}

Together, these lemmas imply the main completeness result given by the
following theorem.

\begin{thm}
\label{thm:completeness}
Let $(G,\pi_0)$ be a colored graph, and let $(G',\pi')$ be its
canonical form.  Then there is a proof that corresponds to $(G,\pi_0)$
deriving the fact $C(G,\pi_0)=(G',\pi')$.\qed
\end{thm}

\section{Implementation and evaluation}

\subsection{Proof format}

The proof is generated by writing the applied rules into a file.
Each rule is encoded as a sequence of numbers that contains the parameters
of the rule sufficient to reconstruct the rule within the
proof-checker. The exact sequences for each of the rules are given in
Table~\ref{table:proof_encoding}, using the notation consistent with the one used in the rules' definitions.

\begin{table}[h!]
\begin{center}
{\footnotesize
\begin{tabular}{|c|p{6.2cm}|}
  \hline 
 \textbf{Rule} & \textbf{Sequence} \\
 \hline\hline 
 \texttt{ColoringAxiom} & $0$ \\
 \hline 
 \texttt{Individualize} & $1,|\nu|,\nu,v,\pi$ \\
 \hline
 \texttt{SplitColoring} & $2,|\nu|,\nu,\pi$ \\
 \hline
 \texttt{Equitable} & $3,|\nu|,\nu,\pi$ \\
 \hline 
 \texttt{TargetCell} & $4,|\nu|,\nu,\pi$ \\
 \hline 
 \texttt{InvariantAxiom} & $5,|\nu|,\nu$ \\
 \hline
 \texttt{InvariantsEqual} & $6,|\nu'|+1,[\nu',v'],\pi_1,|\nu''|+1,[\nu'',v''],\pi_2$ \\
 \hline
 \texttt{InvariantsEqualSym} & $7,|\nu'|,\nu',|\nu''|,\nu''$ \\
 \hline 
 \texttt{OrbitsAxiom} & $8,v,|\nu|,\nu$ \\
 \hline
 \texttt{MergeOrbits} & $9,|\Omega_1|,\Omega_1,|\Omega_2|,\Omega_2,|\nu|,\nu,\sigma,w_1,w_2$ \\
 \hline
 \texttt{PruneInvariant} & $10,|\nu'|+1,[\nu',v'],\pi_1,|\nu''|+1,[\nu'',v''],\pi_2$ \\
 \hline
 \texttt{PruneLeaf} & $11,|\nu'|,\nu',\pi_1,|\nu''|,\nu'',\pi_2$ \\
\hline 
\texttt{PruneAutomorphism} & $12,|\nu'|,\nu',|\nu''|,\nu'',\sigma$ \\
\hline
\texttt{PruneParent} & $13,|\nu|,\nu,|W|,W$ \\
\hline
\texttt{PruneOrbits} & $14,|\Omega|,\Omega,|\nu|,\nu,w_1,w_2$ \\
\hline
\texttt{PathAxiom} & $15$ \\
\hline 
\texttt{ExtendPath} & $16,|\nu|,\nu,|W|,W,w$ \\
\hline
\texttt{CanonicalLeaf} & $17,|\nu|,\nu,\pi$ \\
\hline
 \end{tabular}
 }
 \end{center}
\caption{Number sequences used to encode the rules}
 \label{table:proof_encoding}
\end{table}

Note that each sequence starts with a \emph{rule code}, that is, a
number that uniquely determines the type of the encoded rule.  Subsequent numbers
in the sequence encode the parameters specific for the rule. Vertex
sequences are encoded such that the length of the sequence precedes
the members of the sequence. Vertices are encoded in a zero-based
fashion, i.e.~the vertex $i$ is encoded with the number $i-1$. Sets of
vertices (orbits and cells) are encoded in the same way as vertex
sequences, but is additionally required that the vertices are listed
in the increasing order. Colorings and permutations are encoded as
sequences of values assigned to vertices $1,2,\ldots,n$ in that order
(the colors are, like vertices, encoded in a zero-based fashion).

The whole proof is, therefore, represented by a sequence of numbers,
where the first number is the number of vertices of the graph, followed by the sequences of numbers encoding the applied rules, in the order of their application. When such a sequence
is written to the output file,
each number is considered as a 32-bit unsigned integer, and encoded as
a six-byte UTF-8
sequence\footnote{\url{https://datatracker.ietf.org/doc/html/rfc2279}}
for compactness. Thus, the proof format is not human-readable.

\subsection{Implementation details}
\label{subsec:impl}

For the purpose of evaluation, we have implemented a prototypical
proof checker in the \texttt{C++} programming language. The
implementation contains about 1600 lines of code (excluding the code
for printing debug messages), but most of the code is quite simple and
can be easily verified and trusted. The main challenge was to provide
an efficient implementation of the derived facts database, since the
checker must know which facts are already derived in order to check
whether a rule application is justified. The facts are, just like
rules, encoded and stored in the memory as sequences of numbers. The
checker offers support for two different implementations of the facts
database. The first uses \texttt{C++} standard \texttt{unordered\_set}
container to store the sequences encoding the facts. This
implementation is easier to trust, but is less memory
efficient. Another implementation is based on \emph{radix tries}. This
implementation uses significantly less memory, since the sequences
that encode facts tend to have common prefixes. On the other hand,
since it is an \emph{in-house} solution, it may be considered harder
to trust, especially if we take into account that it is the most
complex part of the checker. However, our intensive code inspecting
and testing have not shown any bugs so far.  The implementation of
radix tries contains about 150 additional lines of code.

We have also implemented \texttt{morphi} --- a prototypical
\texttt{C++} implementation of the canonical form search algorithm,
based on the McKay and Piperno's scheme, instantiated with the concrete
functions $\bar{R}$, $\bar{T}$ and $\bar{\phi}$, as described in this
paper. It is extended with the ability to produce proofs in the
previously described format. The algorithm supports two strategies of
proof generation.

In the first strategy, the proof is generated \emph{during the
  search}, with rules being exported whenever the respective steps of
the algorithm are executed. The proof generated in this way
essentially represents a formalized step-by-step reproduction of the
execution of the search algorithm.

In the second strategy, called the \emph{post-search} strategy, the
rules are exported after the search algorithm has finished and
produced a canonical leaf and a set of generators of the automorphism
group. The proof is then generated by initiating the search tree
traversal once more, this time being able to utilize the pruning
operations more extensively since the knowledge of the canonical leaf
and discovered automorphisms is available from the start.

The proof checker implementation is much simpler than the
implementation of the canonical form search algorithm. Namely, the
canonical search implementation uses many highly optimized
data-structures and algorithms, while almost all proof checker data
structures and algorithms are quite straightforward (usually
brute-force). For example, our canonical search implementation
includes an efficient data structure for representing colorings, a
union-find data-structure for representing orbits, an efficient
data-structure for representing sets of permutations (and finding
permutations that stabilize a given vertex set). It adapts to the
graph size by using different integer types (8-bit, 16-bit or 32-bit).
It uses a specialized memory allocation system (a stack based
memory-pool). The refinement algorithm is specialized in cases when
cells have 1 or 2 elements. The invariant is calculated incrementally
(based on its value in the previous node). All those (and many more)
implementation techniques are absent in the proof checker. If added,
proof checking would become much more complex and harder to implement
and verify.  Since our experiments (see Subsection~\ref{sec:experimental}) show that even with the simplest
implementation the proof checker is not much slower than the canonical
form search, we opted to keep the proof checker implementation as
simple as possible (and much simpler than the original algorithm).

\subsection{Experimental evaluation}
\label{sec:experimental}

The graph instances used for evaluation of the approach are taken
from the benchmark library provided by McKay and
Piperno\footnote{\url{https://pallini.di.uniroma1.it/Graphs.html}},
given in DIMACS format. We included all the instances from the
library in our evaluation, except those instances whose initial
coloring is not trivial (this is because our implementation currently
does not support colored graphs as inputs). In total, our benchmark
set consists of 1284 instances. The experiments were run on a computer
with four AMD Opteron 6168 1.6GHz 12-core processors (that is, 48
cores in total), and with 94GB of RAM.

The evaluation is done using \texttt{morphi}. The first goal of the
evaluation was to estimate how our implementation compares to the
state-of-the-art isomorphism checking implementations. For this
reason, we compared it to the state-of-the-art solver \texttt{nauty}
(also based on the McKay's general scheme for canonical labelling) on
the same set of instances. Both solvers were executed on all 1284
instances, with 60 seconds time limit per instance. For the sake of
fair comparison, the proof generating capabilities were disabled in
our solver during this stage of the evaluation.

\begin{table}[!h]
 \begin{tabular}{|c|c|c|c|}
 \hline
   \textbf{Solver} & \textbf{\# solved} & \textbf{Avg.~time} & \textbf{Avg.~time on solved} \\
 \hline\hline
  \texttt{nauty} & 1170  & 15.48 & 5.3 \\
  \hline
  \texttt{morphi} & 709 & 58.05 & 7.81 \\
  \hline
 \end{tabular}

 \caption{Results of evaluation of \texttt{nauty} and \texttt{morphi}
   on the entire benchmark set. The total number of instances in the
   set was 1284. Times are given in seconds. Time limit per instance
   was 60s. When the average times were calculated, 120s was assumed for unsolved instances.}
 \label{table:nauty}
\end{table}

\begin{figure}[h!]
\begin{center}
 \includegraphics[width=180pt]{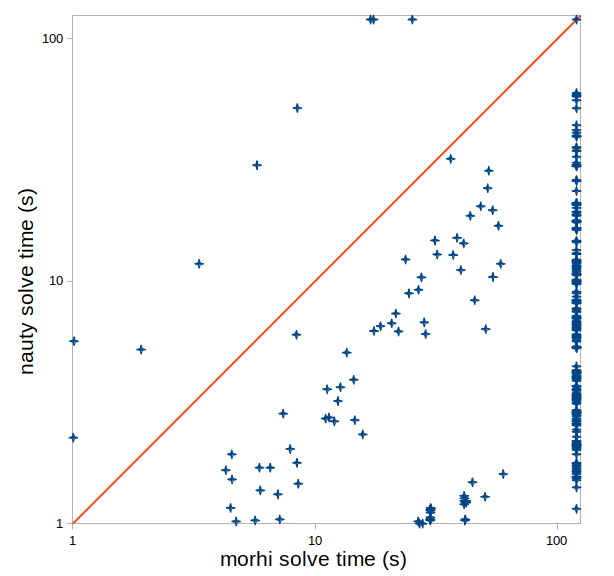}
\end{center}
\caption{A comparison of behaviour of \texttt{nauty} and \texttt{morphi} on particular instances. Times are given in seconds}
\label{fig:plot_nauty}
\end{figure}

The results are given in Table~\ref{table:nauty}. Note that when the
average solving times were calculated, twice the time limit was used
for unsolved instances (that is, 120 seconds). This is the PAR-2
score, that is often used in SAT competitions. The results show that
\texttt{nauty} is significantly faster on average, and it managed to
solve 461 more instances in the given time limit. However, on solved
instances, the average solving times are much closer to each other,
which suggests that our solver is comparable to \texttt{nauty} on a
large portion of the benchmark set (at least on those 709 instances
that our solver managed to solve in the given time limit). A more
detailed, per-instance based comparison is given in Figure~\ref{fig:plot_nauty}. The figure shows that, although most of the
instances are solved faster by \texttt{nauty}, on a significant
portion of them our solver \texttt{morphi} is still comparable (less
than an order of magnitude slower), and there are several instances on
which \texttt{morphi} performed better than \texttt{nauty}. Overall,
we may say that \texttt{morphi} is comparable to the state-of-the-art
solver \texttt{nauty} on average. This is very important, since it
confirms that our prototypical implementation did not diverge too much
from the modern efficient implementations of the McKay's algorithm,
making the approach presented in this paper relevant.

In the rest of this section, we provide the evaluation of the proof
generation and certification, using our solver \texttt{morphi} (with
the proving capabilities enabled), and our prototypical checker, also
described in the previous section\footnote{The implementation of the
  search algorithm and the checker, together with the per-instance
  evaluation details, is available at:
  \url{https://github.com/milanbankovic/isocert}}.  We consider two
different versions of \texttt{morphi}, implementing the two different
proof generating strategies, explained in the previous section:

\begin{itemize}
 \item \texttt{morphi-d}: the \emph{during-search} strategy
 \item \texttt{morphi-p}: the \emph{post-search} strategy
\end{itemize}

The instances used in this stage of evaluation are exactly those
instances that our solver \texttt{morphi} without proof generating
capabilities managed to solve in the time limit of 60 seconds
(i.e.~the instances selected in the previous stage of evaluation, 709
instances in total). Both versions of the solver \texttt{morphi} were
run on all these 709 instances, this time without a time limit, with
the proof generation enabled. In both cases, the generated proofs were
verified using our prototypical checker.

The main result of the evaluation is that the proof verification was
successful for all tested instances, i.e.~for all generated proofs our
checker confirmed their correctness.

For each instance we also measure several parameters that are
important for estimating the impact of the proof generation and
checking on the overall performance, as well as for comparing the two
variants of the proof generating strategy. The \emph{solve time} is
the time spent in computing the canonical form. The \emph{prove time}
is the time spent in the proof generation.  Notice that in case of
\texttt{morphi-d}, we can only measure the sum of the solve time and
the prove time, since the two phases are intermixed. The \emph{check
  time} is the time spent in verifying the proof. We also measure the
\emph{proof size}.

\begin{figure}[h!]
\begin{center}
 \includegraphics[width=430pt]{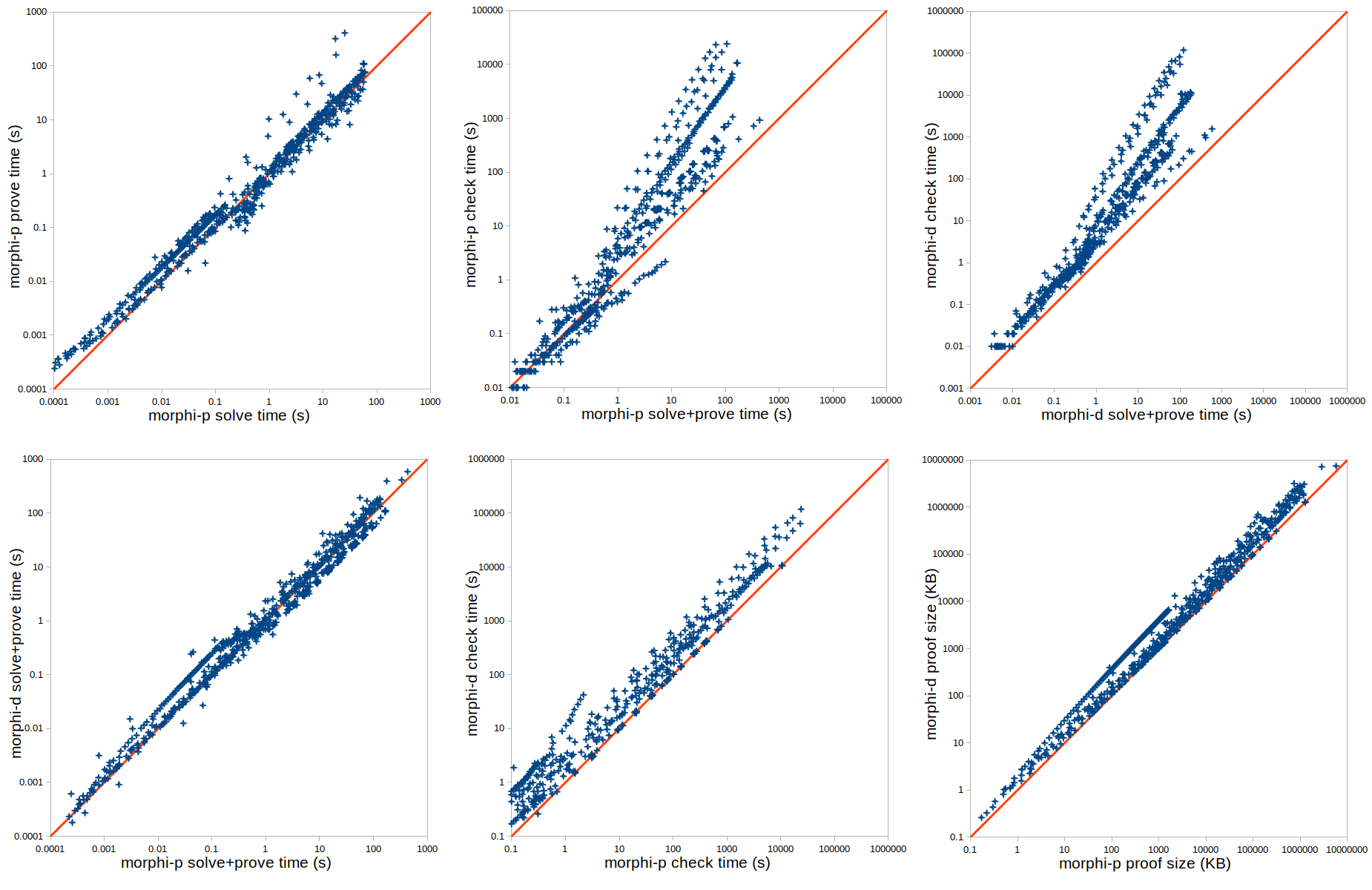}
\end{center}
\caption{Behavior of \texttt{morphi} and the checker on particular
  instances. Times are given in seconds, and proof sizes are given in
  kilobytes}
\label{fig:plots}
\end{figure}

Some interesting relations between the measured parameters on
particular instances are depicted by scatter plots given in
Figure~\ref{fig:plots}, and corresponding minimal, maximal and average
ratios are given in Table~\ref{table:avgs}.  We discuss these
relations in more details in the following paragraphs.

\begin{table}[h!]
\begin{center}
 \begin{tabular}{|l|c|c|c|}
  \hline 
    \textbf{Ratio}      & \textbf{Min} & \textbf{Max} & \textbf{Average} \\
  \hline\hline
  solve/prove (\texttt{morphi-p}) &  0.053 & 4.090 & 0.886 \\
 \hline
 (prove+solve)/check (\texttt{morphi-p}) & 0.003 & 3.547
  & 0.558 \\
 \hline 
 (prove+solve)/check (\texttt{morphi-d}) & 0.001 & 1.009
 & 0.219 \\
 \hline 
 \texttt{morphi-p}/\texttt{morphi-d} (solve+prove) & 0.169 & 2.529
 & 1.070 \\
 \hline 
 \texttt{morphi-p}/\texttt{morphi-d} (check) & 0.048 & 1.192
 & 0.555 \\
 \hline 
 \texttt{morphi-p}/\texttt{morphi-d} (proof size) & 0.166
 & 1.000 & 0.674 \\
 \hline 
 \end{tabular}
\end{center}
\caption{Minimum, maximum and average ratios between the observed parameters on all tested instances}
 \label{table:avgs}
\end{table}

\paragraph{Solve time vs. prove time.}
Top-left plot in Figure~\ref{fig:plots} shows the relation between the
solve time and the prove time for \texttt{morphi-p}. The solve time
was about 89\% of the prove time on average, which means that the
proof generation tends to take slightly more time than the search. In
other words, the invocation of the algorithm with proof generation
enabled on average takes over twice as much time as the canonical form
search alone. This was somewhat expected, since the proof is generated
by performing the search once more, using the information gathered
during the first tree traversal.

\paragraph{Solve+prove time vs. check time.}
Top-middle and top-right plots in Figure~\ref{fig:plots} show the
relation between solve+prove time and check time, for
\texttt{morphi-p} and \texttt{morphi-d}, respectively. From the plots
it is clear that the proof verification tends to take significantly
more time than the search and the proof generation together, and this
is especially noticeable on ``harder'' instances. The average
(solve+prove)/check ratio was 0.558 for \texttt{morphi-p}, with
minimal ratio on all instances being 0.003. The phenomenon is even
more prominent for \texttt{morphi-d} (0.219 average, 0.001
minimal). Such behavior can be explained by the fact that the checker
is implemented with the simplicity of the code in mind (in order to
make it more reliable), and not its efficiency. Most of the conditions
in the rule applications are checked by brute force. On the other
hand, the search algorithm's implementation is heavily optimized for
best performance.

\paragraph{Comparing the proof generating strategies.}
Bottom-left, bottom-middle and bottom-right plots in
Figure~\ref{fig:plots} compare the \texttt{morphi-p} and
\texttt{morphi-d} with respect to the solve+prove times, check times
and proof sizes, respectively.  The solve+prove times were similar on
average (\texttt{morphi-d} being slightly faster, with the average
solve+prove time equal to 17.7s, while for \texttt{morphi-p} it was
18.9s).  Most of the points on the bottom-left plot are very close to
the diagonal, almost evenly distributed on both sides of it.  The
average solve+prove time ratio was very close to 1 (1.07).  This
suggests that, when the time consumed by the algorithm is concerned,
it does not matter which proof generation strategy is employed.  On
the other hand, the size of a \texttt{morphi-p} proof was about 67\%
of the size of the corresponding \texttt{morphi-d} proof on average,
and a \texttt{morphi-p} proof was never greater than the corresponding
\texttt{morphi-d} proof.  Finally, the check time for proofs generated
by \texttt{morphi-p} was about 55.5\% of the check time for proofs
generated by \texttt{morphi-d} on average. This means that the proof
checking will be more efficient if we employ the post-search proof
generation strategy. One of the main reasons for that is significantly
smaller average proof size, but this might not be the only
reason. Namely, the check time does not depend only on the proof size,
but also on the structure of the proof, since some rules are harder to
check than others. The most expensive check is required by
\texttt{Equitable}, \texttt{InvariantsEqual}, \texttt{PruneInvariant}
and \texttt{PruneLeaf} rules, since it includes time consuming
operations such as verifying the equitability of the coloring,
calculating the hash function for colored graphs, or comparing the
graphs. These operations tend to be more expensive as we move towards
the leaves of the search tree, i.e.~as the colorings become closer to
discrete. In post-search proof generating strategy, the prunings tend
to happen higher in the search tree during the second tree traversal
(when the proof is actually generated), lowering the proportion of
these expensive rules in the proof.

\subsection{False proofs}

It is important to stress that the checker does not only accept
correct proofs, but that it indeed rejects false proofs as well. Some
issues with the solver were found and resolved during development as a
consequence of false proofs that had been rejected by the checker. For
example, there were issues with the invariant calculation and the
coloring refinement that led to incorrect canonical forms and proofs
being produced for some instances. There was also an issue that caused
emission of false facts into the proof in some occasions, specifically
during the traversal of a pruned subtree in search for automorphisms.
As a part of further testing, several artificial bugs were introduced
into the solver which were successfully caught by the checker.%
\footnote{Some of the artificial bugs that we introduced for testing
are provided along with the source code, so the interested reader may
apply them to see how the checker responds to such false proofs.}

\section{Related work}

\paragraph{Certifying algorithms in general.}

There are many applications in the industry or science where a user
may benefit of having an algorithm which produces certificates that
can confirm the correctness of its outputs. McConnell et
al.~\cite{mcconnell} go one step further, suggesting that \emph{every}
algorithm should be \emph{certifying}, i.e.~be able to produce
certificates. They establish a general theoretical framework for
studying and developing certifying algorithms, and provide a number of
examples of practical algorithms that could be extended to produce
certificates. The authors stress two important properties of a
certifying algorithm that must be fulfilled: \emph{simplicity}, which
means that it is easy to prove that the (correct) certificates really
certify the correctness of the obtained outputs, and
\emph{checkability}, which means that it is easy to check the
correctness of a certificate. In both cases, the property of
\emph{being easy} is not strictly defined.

In case of simplicity, it is expected that the fact that a certificate
really proves the correctness of the corresponding output should be
obvious and easily understandable. This may be considered as a
potential drawback of our approach, since the soundness of our proof
system is not obvious and requires a non-trivial proof. However, we
did our best to provide an as precise as possible soundness proof in
the appendix of this paper, with hope that it is convincing enough to
the reader (we also prove the completeness, i.e.~the existence of a
certificate for each input, for which McConnell et
al.~\cite{mcconnell} do not insist to be simple to prove). We have also
formalized the soundness proof within Isabelle/HOL theorem prover,
and we believe that the existence of a formal soundness proof is a
good substitution for the simplicity property stressed by the authors.

When checkability is concerned, the authors suggest that an algorithm
for certificate checking should either have a simple logical
structure, such that its correctness can be easily established, or it
should be formally verified \cite{mcconnell}.  We believe that our
checker indeed has a simple logical structure, since its most
complicated parts such as calculating $\operatorname{split}(G,\pi,i)$
function are implemented using a brute force, without any sofisticated
data structures or programming techniques. Moreover, we have already
argued (see Subsection~\ref{subsec:impl}) that its implementation is much
simpler than the implementation of the canonical labelling algorithm
itself, and can be more easily verified. In fact, we have already
provided and formally verified an abstract proof checker specification
in Isabelle/HOL. The remaining step is to refine this abstract
specification into an efficient executable code, which we plan
to do in the future.

McConnell et al.~\cite{mcconnell} also suggest that the complexity of
a certificate checking algorithm should preferably be linear with
respect to the size of its input, and its input is composed of an
input/output pair of the certifying algorithm, and of the
corresponding certificate. In our case, each rule from the certificate
is read and checked exactly once, but checking of some rules may
require traversing the graph's adjacency matrix, so the complexity is
not exactly linear. However, since the size of the proof is usually
much greater than the size of the graph, the complexity may be
considered to be almost linear.

\paragraph{Certifying algorithms in SAT solving.}

Certificate checking has been successfully used, for example, in
certifying unsatisfiability results given by SAT or SMT solvers
\cite{drat-trim,largest-proof,lammich_unsat}, and this enabled
application of external SAT and SMT solvers from within interactive
theorem provers \cite{sat_isabelle,bohme_phd,sledgehammer-smt}.

The proof system used in the case of SAT solvers is very simple and is
based on the resolution. While early approaches considered simple
resolution proofs \cite{zhang2003validating} which were very easy to
check, but much harder to produce, modern approaches are mostly based
on so-called \emph{reversed unit-propagation} (RUP) proofs
\cite{Gelder2008}, which are easy to generate, but harder to
check. While this fact violates the checkability property to some
extent, it is widely adopted by the SAT solving community, and proof
formats such as DRAT \cite{drat-trim} became an industry standard in
the field. Our proof system has a similar trait, since we retained
some complex computations within the checker, in order to make our
proofs smaller and easier to generate. As with RUP proofs, we consider
this as a good tradeoff.

\paragraph{Certifying pseudo-boolean reasoning.}
The SAT problem is naturally generalized to the \emph{pseudo-boolean}
(PB) satisfiability problem \cite{roussel2021pseudo}, where it is also
useful to have certifying capabilities. Recently, a simple but
expressive proof format for proving the PB unsatisfiability was
proposed by Elffers et al.~\cite{elffers2020justifying}, based on
\emph{cutting planes} \cite{cook1987complexity}. That proof format was
then employed for expressing unsatisfiability proofs for different
combinatorial search and optimization problems, such as constraint
satisfaction problems involving the alldifferent constraint
\cite{elffers2020justifying}, the subgraph isomorphism problem
\cite{gocht2021subgraph}, and the maximum clique and the maximum
common subgraph problem \cite{gocht2020certifying}. The main idea is
to express some combinatorial problem $P$ as a pseudo-boolean problem
$PB(P)$, and then to express the reasonings of a dedicated solver for
$P$ in terms of PB constraints implied by the constraints of $PB(P)$
(such PB constraints are exported by the dedicated solver during the
solving process).  The obtained PB proof is then checked by an
independent checker which relies on a small number of simple rules
(such as addition of two PB constraints or multiplying/dividing a PB
constraint with a positive constant), as well as on generalized RUP
derivations \cite{elffers2020justifying}. The authors advocate the
generality of their approach, enabling it to express the reasoning
employed in solving many different combinatorial problems, using a
simple and general language which does not know anything about the
concepts and specific semantics of considered problems.

In contrast, our proof system is tied to a particular problem -- the
problem of constructing canonical graph labellings, and a particular
class of algorithms -- those that could be instantiated from the
McKay/Piperno scheme \cite{mckay_piperno}.\footnote{Actually, our
  proof system is tied to a particular instance of McKay/Piperno's
  scheme, fixed by a choice of $R$, $T$ and $\phi$ functions, but it
  should be relatively easy to adapt it for other such
  instantiations.}  Citing McKay, the canonical forms returned by the
algorithm considered in this paper are ``designed to be efficiently
computed, rather than easily described'' \cite{mckay_isomorph_free}.
Since the canonical labelling specification is given in terms of the
algorithm, it seems that certification must also be tied to it. This
is the main reason that led to our decision to develop an
algorithm-specific proof format which is powerful enough to capture
all the reasonings that appear in the context of this particular class
of labelling algorithms.

The question remains whether the canonical labelling problem, like
some other graph problems previously mentioned, may be encoded in
terms of PB constraints in a way simple enough to be trusted (as noted
by Gocht et al.~\cite{gocht2020certifying}, an encoding should be
simple -- since it is not formally verified, any errors in the
encoding may lead to ``proving the wrong thing''). Moreover, the
encoding should permit expressing the operations made by the canonical
labelling algorithm in terms of PB constraints implied by the
encoding. Up to the authors' knowledge, it is not obvious whether such
encoding exists. The main issue that we see in such an approach is
that the canonical form of the graph, as defined by McKay and Piperno
\cite{mckay_piperno}, is not simple to describe in advance using PB
constraints, since it is defined implicitly by the canonical
labelling algorithm itself. For instance, one should capture the
notion of ``coarsest equitable coloring finer than a given coloring
$\pi$'', which is especially hard if we know that such a coloring is
not unique, so we must fix one particular ordering of the cells (and
that ordering must be also captured using the PB language). The
similar case is with describing hash functions used in calculating
node invariants.

\section{Conclusions and further work}

We have formulated a proof-system for certifying that a graph is a
canonical form of another graph, which is a key step in verifying
graph isomorphism. The proof system is based on the state-of-the-art
scheme of McKay and Piperno and we proved its soundness and
completeness. Our algorithm implementation produces proofs that can be
checked by an independent proof checker. The implementation of the
proof checker is much simpler than the implementation of the core
algorithm, so it is easier to trust.

One of the main problems with our approach is that our proof format is
very closely related to the canonical labelling algorithm
itself. Therefore, it is very difficult to use our proof format and
proof checker for different approaches to isomorphism checking. On the
other hand, the McKay/Piperno scheme is broad enough to cover several
state-of-the-art algorithms, and we believe that our proofs and
checkers can be easily adapted to certify all these algorithms. Our
prototype implementation \texttt{morphi} is slower compared to
state-of-the-art implementations such as \texttt{nauty}, but still
comparable on many instances. We plan to make further optimizations to
\texttt{morphi}. We are also considering adapting \texttt{nauty}'s
source-code to emit proofs in our format, which is not trivial, as
\texttt{nauty} has been developed for more than 30 years and is a very
complex and highly optimized software system.

The main direction of our further work is to go one step further and
fully integrate isomorphism checking into an interactive theorem
prover. We have already taken important steps in this direction: we
have formally defined the McKay and Piperno's abstract scheme in
Isabelle/HOL and formally proved its correctness (Theorem~\ref{thm:abstract_correctness}), we have shown that our concrete
functions $\bar{R}$, $\bar{T}$, and $\bar{\phi}$ satisfy the
McKay/Piperno axioms, and we have formalized the rules of our proof
system and formally proved its soundness. We have also provided and
formally verified an abstract
proof checker specification in Isabelle/HOL. In our further work it
remains to refine our abstract proof checker specification into
efficient executable code, which we plan to do using Imperative
Refinement Framework \cite{Refine_Imperative_HOL-AFP}.

Another direction of our further work is to examine the
possibilities of reducing the complexity of our proof system.
Currently, our system operates with 9 different types of facts and has 18 different
rules. Compared to some other well-known proof systems, such as those
used in SAT or PB solving, which rely on few very simple rules,
our system may be much harder to understand and trust, and proof
checking becomes a non-trivial task. We tried to fill this gap by
providing a formal proof of the soundness of our system in
Isabelle/HOL, together with a (verified) abstract specification of
our proof checker. However, it would be beneficial to develop a simpler
proof system or to find a way to reduce our system to some existing
proof system which is well understood and trusted (such as those
mentioned above). We are currently not aware if this is possible in
case of graph canonical labelling algorithms based on McKay and Piperno's
abstract scheme, so this remains a great challenge that may be
interesting to tackle in further work.

\section*{Acknowledgment}
  \noindent This work was partially supported by the Serbian Ministry of
  Science grant 174021. We are very grateful to the anonymous reviewers
  whose insightful comments and remarks helped us to make this text much
  better.

\bibliographystyle{alphaurl}
\bibliography{rules}

\appendix

\section{Proofs of lemmas and theorems}

\theoremstyle{thmC}\newtheorem*{repthm}{Theorem\hspace*{-4pt}}
\theoremstyle{thmC}\newtheorem*{replem}{Lemma\hspace*{-4pt}}

\begin{repthm}[\textbf{\ref{thm:abstract_correctness}}]
  If $R$, $T$, and $\phi$ satisfy all axioms, then for the function $C$
  determined by $R$, $T$ and $\phi$ it holds that two colored graphs
  $(G, \pi_0)$ and $(G', \pi_0')$ are isomorphic if and only if
  $C(G, \pi_0) = C(G', \pi_0')$.
\end{repthm}

\begin{proof}
  Assume that $(G, \pi_0)$ and $(G', \pi_0')$ are isomorphic. Then
  exists a permutation $\sigma \in S_n$ such that
  $(G',\pi_0') = (G^\sigma, \pi_0^\sigma)$.  Since the functions $R$
  and $T$ are label-invariant, it is easy to show that
  $\mathcal{T}(G^\sigma, \pi_0^\sigma, \nu^\sigma) =
  \mathcal{T}(G,\pi_0,\nu)^\sigma$. Moreover, since $\phi$ is also
  label-invariant, if $\nu$ is a leaf from $\mathcal{T}(G,\pi_0)$, and
  $\nu^\sigma$ is the corresponding leaf from
  $\mathcal{T}(G^\sigma,\pi_0^\sigma)$, then
  $\phi(G,\pi_0,\nu) = \phi(G^\sigma,\pi_0^\sigma,\nu^\sigma)$, so the
  leaves that correspond to each other with respect to $\sigma$ have
  equal node invariants. Therefore, the sets of maximal leaves of the
  two trees correspond to each other, with respect to
  $\sigma$. Furthermore, if $R(G,\pi_0,\nu) = \pi$, for some leaf
  $\nu \in \mathcal{T}(G, \pi_0)$, then
  $R(G^\sigma, \pi_0^\sigma, \nu^\sigma)=R(G,\pi_0,\nu)^\sigma =
  \pi^\sigma = \sigma^{-1}\pi$, and we have
  $(G^\sigma)^{\pi^\sigma} = (G^\sigma)^{\sigma^{-1}\pi} =
  G^{\sigma\sigma^{-1}\pi} = G^\pi$, so the graphs that correspond to
  the leafs $\nu \in \mathcal{T}(G,\pi_0)$ and
  $\nu^\sigma \in \mathcal{T}(G^\sigma, \pi_0^\sigma)$ are
  identical. Therefore, the maximal graphs that correspond to the
  maximal leaves in both trees are also identical, so
  $C(G,\pi_0) = C(G^\sigma, \pi_0^\sigma) = C(G',\pi_0')$ (although
  the canonical leaves of the trees $\mathcal{T}(G,\pi_0)$ and
  $\mathcal{T}(G^\sigma, \pi_0^\sigma)$ do not have to correspond to
  each other with respect to $\sigma$).
  
  Conversely, assume that $C(G,\pi_0)=C(G',\pi_0')$. Since
  $C(G,\pi_0)$ is a graph assigned to some leaf of
  $\mathcal{T}(G,\pi_0)$, then there exists a permutation
  $\sigma \in S_n$ such that $C(G,\pi_0) = (G^\sigma, \pi_0^\sigma)$.
  Similarly, there exists a permutation $\sigma' \in S_n$ such that
  $C(G',\pi_0') = (G'^{\sigma'}, \pi_0'^{\sigma'})$. Since
  $(G^\sigma, \pi_0^\sigma)=(G'^{\sigma'}, \pi_0'^{\sigma'})$, we have
  $(G',\pi_0')=(G^\sigma,\pi_0^\sigma)^{\sigma'^{-1}} =
  (G,\pi_0)^{\sigma\sigma'^{-1}}$, so $(G,\pi_0)$ and $(G',\pi_0')$
  are isomorphic.
\end{proof}

\begin{replem}[\textbf{\ref{lemma:make_equitable_1}}]
  For each $\sigma \in S_n$,
  $\mathtt{make\_equitable}(G^\sigma, \pi^\sigma, \alpha^\sigma) =
  \mathtt{make\_equitable}(G,\pi,\alpha)^\sigma$.
\end{replem}

\begin{proof}
  Let $\pi'$ and $\pi_\sigma'$ be the current colorings in invocations
  $\mathtt{make\_equitable}(G,\pi,\alpha)$ and
  $\mathtt{make\_equitable}(G^\sigma, \pi^\sigma, \alpha^\sigma)$
  respectively. Similarly, let $\alpha$ and $\alpha_\sigma$ be the
  current lists of splitting cells in the two invocations. We prove by
  induction that, for each $k$, after $k$ iterations of the $while$
  loop in both invocations it holds that $\pi_\sigma' = {\pi'}^\sigma$
  and $\alpha_\sigma = \alpha^\sigma$. The statement holds initially,
  for $k=0$, since $\pi'=\pi$, $\pi_\sigma' = \pi^\sigma$ and
  $\alpha_\sigma = \alpha^\sigma$. Assume that statement holds for
  some $k$. First notice that $\pi'$ is a discrete coloring after $k$
  iterations if and only if $\pi_\sigma'$ is discrete, since
  $\pi_\sigma' = {\pi'}^\sigma$, by induction hypothesis. Similarly,
  $\alpha$ is empty if and only if $\alpha_\sigma$ is empty, since
  $\alpha_\sigma = \alpha^\sigma$. Therefore, either both invocations
  enter the next iteration of the $while$ loop, or both finish the
  execution of the algorithm after $k$-th iteration, returning $\pi'$
  and $\pi'^\sigma$, respectively.  In the first case, if $W$ is the
  first cell of $\pi'$ that belongs to $\alpha$, then $W^\sigma$ is
  the first cell of $\pi_\sigma'$ belonging to $\alpha_\sigma$, since
  $\pi_\sigma' = {\pi'}^\sigma$ and $\alpha_\sigma = \alpha^\sigma$ by
  induction hypothesis, and the relabelling does not change the order
  of the cells. Moreover, if a cell $X \in \pi'$ is partitioned into
  $X_1,X_2,\ldots,X_k$, then the corresponding cell
  $X^\sigma \in {\pi'}^\sigma$ is partitioned into
  $X_1^\sigma,X_2^\sigma,\ldots,X_k^\sigma$, with
  $|X_i^\sigma| = |X_i|$, since the partitioning (and the order of
  partitions) depends only on the number of adjacent vertices from
  $W$, which is not affected by relabelling. This implies that the
  statement also holds after $k+1$ iterations of the $while$
  loop. This proves the lemma, since both invocations finish after
  equal numbers of iterations.
\end{proof}

\begin{replem}[\textbf{\ref{lemma:make_equitable_2}}]
  If $\pi' = \mathtt{make\_equitable}(G,\pi,\alpha)$, then
  $\pi' \preceq \pi$.
\end{replem}

\begin{proof}
  The lemma follows from the fact that the only type of transformation
  that is applied to the given coloring $\pi$ by the procedure
  $\mathtt{make\_equitable}$ is splitting of its cells, which always
  produces a finer coloring.
\end{proof}

\begin{replem}[\textbf{\ref{lemma:R_correctness}}]
  If $\pi = \bar{R}(G,\pi_0,\nu)$, then:
  \begin{enumerate}
   \item $\pi \preceq \pi_0$
   \item $\pi$ individualizes vertices from $\nu$
   \item for each $\sigma \in S_n$,
     $\bar{R}(G^\sigma,\pi_0^\sigma,\nu^\sigma) =
     \bar{R}(G,\pi_0,\nu)^\sigma$
   \end{enumerate}
\end{replem}
\begin{proof}
  The lemma will be proved by induction on $k = |\nu|$. For $k=0$, we
  have $\nu = [\ ]$, the statement follows from
  Lemma~\ref{lemma:make_equitable_1} and
  Lemma~\ref{lemma:make_equitable_2}.  Assume that the statement holds
  for some $k$ and let $\nu = [\nu',v]$, where $|\nu'|=k$. Notice that
  $\bar{R}(G, \pi_0, [\nu',v])$ first invokes
  $\bar{R}(G,\pi_0,\nu')$. The statement holds for the coloring $\pi'$
  returned by the recursive call, by induction hypothesis. The
  procedure additionally individualizes the vertex $v$, obtaining the
  coloring $\pi''$, and then calls
  $\mathtt{make\_equitable}(G,\pi'',\alpha)$, where
  $\alpha = \{ \{ v \} \}$. Let $\pi$ be the coloring returned by this
  call. Since $\pi''$ is obtained from $\pi'$ by individualization of
  the vertex $v$, it holds $\pi'' \preceq \pi'$.  By
  Lemma~\ref{lemma:make_equitable_2}, we have $\pi \preceq \pi''$, and
  by induction hypothesis, it holds $\pi' \preceq \pi_0$. Therefore,
  $\pi \preceq \pi_0$, by transitivity of $\preceq$. Since $\pi'$
  individualizes the vertices from $\nu'$ (by induction hypothesis),
  then coloring $\pi$ individualizes the vertices from $\nu$ (since
  $\pi \preceq \pi'' \preceq \pi'$ and the vertex $v$ is also
  individualized in $\pi''$). Finally, since
  $\bar{R}(G^\sigma,\pi_0^\sigma,\nu'^\sigma) = \pi'^\sigma$ (by
  induction hypothesis), and since the cell containing $v^\sigma$ in
  $\pi'^\sigma$ is $W^\sigma$, where $W$ is the cell containing $v$ in
  $\pi'$, it follows that the coloring obtained from $\pi'^\sigma$
  after the individualization of $v^\sigma$ will be equal to
  $\pi''^\sigma$, where $\pi''$ is obtained from $\pi'$ by
  individualization of $v$. Furthermore,
  $\{ \{ v^\sigma \} \} = \alpha^\sigma$, so we have
  $\mathtt{make\_equitable}(G^\sigma, \pi''^\sigma, \alpha^\sigma) =
  \mathtt{make\_equitable}(G,\pi'', \alpha)^\sigma$, by
  Lemma~\ref{lemma:make_equitable_1}.
\end{proof}

\begin{replem}[\textbf{\ref{lemma:make_equitable}}]
Let $W$ be the cell of $\pi'$ that is chosen as a splitting cell in
 some arbitrary iteration of the \texttt{while} loop of the procedure
 $\mathtt{make\_equitable}(G, \pi, \alpha)$. Let $W'$ be any cell of
 $\pi'$ that precedes $W$ in $\pi'$ (that is, it corresponds to a
 smaller color). Then $W'$ does not cause any splitting, i.e.,~for any
 two vertices of a same color in $\pi'$ the numbers of their adjacent
 vertices from $W'$ are equal.
\end{replem}
\begin{proof}
  To simplify terminology, we call the cells $W'$ satisfying the
  condition stated in the last sentence of the lemma \emph{neutral}.
  More precisely, we say that a set $X$ of vertices from a graph $G$
  is \emph{neutral} with respect to a coloring $\pi$, if for each cell
  $W$ of $\pi$ and for each two vertices $v_1$ and $v_2$ from $W$ the
  numbers of vertices from $X$ adjacent to $v_1$ and $v_2$ are equal.
  In that terminology, the lemma claims that any cell $W'$ of $\pi'$
  preceding $W$ is neutral with respect to $\pi'$.

  Note that if some set $X$ is neutral with respect to a coloring
  $\pi$, then it is also neutral with respect to any coloring
  $\overline{\pi}$ finer than $\pi$.

  Furthermore, we say that a cell $X$ of $\pi$ is \emph{conditionally
    neutral} with respect to $\pi$ if for each coloring
  $\overline{\pi}$ finer than or equal to $\pi$ if all the cells that
  precede $X$ in $\pi$ are neutral with respect to $\overline{\pi}$,
  then $X$ is also neutral with respect to $\overline{\pi}$. Note that
  if a cell $X$ is neutral with respect to $\pi$, it is also
  conditionally neutral with respect to $\pi$.

  We first prove the following two propositions.

\begin{prop}\label{prop1} Let $X$ be a cell of a coloring
  $\pi$ which is conditionally neutral with respect to $\pi$, and let
  $\overline{\pi}$ be any coloring finer than $\pi$ (i.e.~obtained by
  partitioning some of its cells).  Let $X_1,\ldots,X_k$ (in that
  order) be partitions of $\overline{\pi}$ obtained by partitioning
  the cell $X$ of $\pi$.  Then $X_k$ is conditionally neutral with
  respect to $\overline{\pi}$.
\end{prop}
  To prove this proposition, first notice that
  $\overline{\pi} \preceq \pi$. Let $\hat{\pi}$ be any coloring finer
  than $\overline{\pi}$ such that all the cells of $\overline{\pi}$
  that precede $X_k$ are neutral with respect to $\hat{\pi}$. We must
  prove that $X_k$ is also neutral with respect to $\hat{\pi}$. Let
  $Y$ be any cell of $\pi$ that precedes $X$ in $\pi$, and let
  $Y_1,\ldots,Y_m$ be the cells of $\overline{\pi}$ obtained by
  partitioning the cell $Y$. Since $Y_1,\ldots,Y_m$ precede $X_k$ in
  $\overline{\pi}$, these cells are neutral with respect to
  $\hat{\pi}$ by assumption. Then it is easy to argue that $Y$ is also
  neutral with respect to $\hat{\pi}$ -- the number of adjacent
  vertices from $Y$ to each vertex $v$ is obtained by summing the
  numbers of adjacent vertices from $Y_1,\ldots,Y_m$, so for any two
  same-colored vertices (with respect to $\hat{\pi}$) such sums must
  be equal. Now, since $\hat{\pi} \preceq \overline{\pi} \preceq \pi$,
  and all the cells that precede $X$ in $\pi$ are neutral with respect
  to $\hat{\pi}$, then $X$ must be also neutral with respect to
  $\hat{\pi}$ (because it is conditionally neutral with respect to
  $\pi$ by assumption). Furthermore, since the cells
  $X_1,\ldots,X_{k-1}$ are also neutral with respect to $\hat{\pi}$ by
  assumption (because they precede $X_k$ in $\overline{\pi}$), then
  the cell $X_k$ is also neutral with respect to $\hat{\pi}$ (since
  $X$ is neutral with respect to $\hat{\pi}$, and the number of
  adjacent vertices from $X_k$ to any vertex $v$ can be obtained by
  subtracting the sum of the numbers of its adjacent vertices from
  $X_1,\ldots,X_{k-1}$ from the number of its adjacent vertices from
  $X$. Again, for the same-colored vertices with respect to
  $\hat{\pi}$, such numbers must be equal). Now it follows that $X_k$
  is conditionally neutral with respect to $\overline{\pi}$, as stated
  by the proposition.

\begin{prop}\label{prop2} Let $X_1,\ldots,X_k$ be the
first $k$ cells of a coloring $\pi$. If $X_1,\ldots,X_k$ are
conditionally neutral with respect to $\pi$, then they are also
neutral with respect to $\pi$.
\end{prop}

This proposition may be proven by induction on the cell index.  For
the cell $X_1$ the statement trivially holds, since there are no cells
that precede it in $\pi$, so it must be neutral with respect to any
coloring finer than or equal to $\pi$, including $\pi$ itself. Assume
the statement holds for $X_1,\ldots, X_{i-1}$, and let us prove that
it holds for $X_i$ ($i \le k$).  Since $X_i$ is conditionally neutral
with respect to $\pi$ it must be neutral with respect to any coloring
finer than or equal to $\pi$ such that all the cells
$X_1,\ldots,X_{i-1}$ are also neutral with respect to it. By induction
hypothesis, $X_1,\ldots,X_{i-1}$ are neutral with respect to $\pi$,
and since $\pi \preceq \pi$, the cell $X_i$ must also be neutral with
respect to $\pi$.

\begin{prop}\label{prop3} In the procedure
$\mathtt{make\_equitable}(G, \pi, \alpha)$, at the beginning of each
iteration of the \texttt{while} loop, any cell $X$ of $\pi'$ that is
not present in $\alpha$ is conditionally neutral with respect to
$\pi'$.
\end{prop}

This claim can be proven by induction on the number of iterations. In
the base case, before the first iteration of the loop (when
$\pi' = \pi$), we have two cases to consider. The first case is when
the procedure is called initially, with $\alpha$ containing all the
cells of $\pi$.  In that case, the statement trivially holds, since
there are no cells of $\pi'$ that are not in $\alpha$.  The second
case is when the procedure is invoked with only one singleton cell
$\{ v \}$ in $\alpha$, where $v$ is the last individualized vertex.
Let $\overline{\pi}$ be the coloring obtained from $\pi$ by
partitioning the cell $V$ containing $v$ into $\{ v \}$ and
$V \setminus \{ v \}$ (in that order).  Since the coloring $\pi$ was
already equitable before individualization of $v$, all the cells of
$\pi$ (including $V$) are neutral with respect to $\pi$. Therefore,
all the cells of $\pi$ distinct from $V$ (which are also the cells of
$\overline{\pi}$ and are not in $\alpha$) are neutral with respect to
$\overline{\pi}$, and therefore, conditionally neutral with respect to
$\overline{\pi}$.  On the other hand, since $V$ is neutral with
respect to $\pi$, it is also conditionally neutral with respect to
$\pi$. By \autoref{prop1}, the cell $V \setminus \{ v \}$ is
conditionally neutral with respect to $\overline{\pi}$. Thus, all
cells of $\overline{\pi}$ not in $\alpha$ are conditionally neutral
with respect to $\overline{\pi}$.

In order to prove the induction step, let us assume that the statement
holds before $k$-th iteration. Let $W \in \alpha$ be the next
splitting cell chosen by the algorithm, and let $\overline{\pi}$ be
the coloring obtained from $\pi'$ by partitioning its cells with
respect to $W$. At the end of $k$-th iteration, the following cells of
$\overline{\pi}$ will not be in $\alpha$:

\begin{itemize}
\item since $W$ is removed from $\alpha$ at the beginning of $k$-th
  iteration, if $W$ is partitioned to $W_1,\ldots,W_k$ (in that
  order), then $W_k$ will not be in $\alpha$. Since $\overline{\pi}$
  is obtained from $\pi'$ by partitioning its cells with respect to
  $W$, then $W$ is neutral with respect to $\overline{\pi}$, and also
  with respect to any coloring $\hat{\pi}$ finer than
  $\overline{\pi}$. Therefore, if $W_1,\ldots,W_{k-1}$ are neutral
  with respect to $\hat{\pi}$, the cell $W_k$ will be also neutral
  with respect to $\hat{\pi}$. By definition, it follows that $W_k$ is
  conditionally neutral with respect to $\overline{\pi}$.

\item for any cell $X$ of $\pi'$ that was not in $\alpha$ at the
  beginning of $k$-th iteration, if $X$ is partitioned to
  $X_1,\ldots,X_k$ (in that order), $X_k$ will not be in $\alpha$. By
  induction hypothesis, $X$ is conditionally neutral with respect to
  $\pi'$, so by \autoref{prop1}, $X_k$ is conditionally neutral with
  respect to $\overline{\pi}$.
\end{itemize}

Therefore, all the cells of $\overline{\pi}$ not in $\alpha$ at the
end of $k$-th iteration are conditionally neutral with respect to
$\overline{\pi}$. By proving the induction step, we have proven the
proposition.

Now let us return to the proof of the main lemma. It easily follows
from the previous propositions, since all the cells preceding $W$ in
$\pi'$ are not in $\alpha$, thus conditionally neutral with respect to
$\pi'$ by \autoref{prop3}, and therefore neutral with respect to $\pi'$
by \autoref{prop2}.
\end{proof}

\begin{replem}[\textbf{\ref{lemma:T_label_invariant}}]
  For each $\sigma \in S_n$ it holds
  $\bar{T}(G^\sigma, \pi_0^\sigma, \nu^\sigma) =
  \bar{T}(G,\pi_0,\nu)^\sigma$.
\end{replem}

\begin{proof}
  The lemma holds since, according to Lemma~\ref{lemma:R_correctness},
  $\bar{R}(G^\sigma,\pi_0^\sigma,\nu^\sigma) =
  \bar{R}(G,\pi_0,\nu)^\sigma$, so their first non-singleton cells
  also correspond to each other, with respect to $\sigma$.
\end{proof}

\begin{replem}[\textbf{\ref{lemma:phi_correct}}]
The following properties of the function $\bar{\phi}$ hold:
\begin{enumerate}
\item If $|\nu'| = |\nu''|$ and
  $\bar{\phi}(G,\pi_0,\nu') <
  \bar{\phi}(G,\pi_0,\nu'')$, then
  $\bar{\phi}(G,\pi_0,[\nu', w_1]) <
  \bar{\phi}(G,\pi_0,[\nu'',w_2])$, for each
  $w_1 \in T(G,\pi_0,\nu')$,
  $w_2 \in T(G,\pi_0,\nu'')$
\item For each $\nu$ and nonempty $\nu'$ such that
  $[\nu, \nu']$ is a tree node, it holds that
  $\bar{\phi}(G, \pi_0, \nu) < \bar{\phi}(G, \pi_0, [\nu, \nu'])$.
\item
  $\bar{\phi}(G^\sigma, \pi_0^\sigma, \nu^\sigma) =
  \bar{\phi}(G,\pi_0,\nu)$ for each node $\nu$ and for each permutation
  $\sigma \in S_n$
\end{enumerate}
\end{replem}

\begin{proof}
The first two properties hold trivially, since the node invariants
are compared lexicographically. For the third property, recall that we assumed that the function $\operatorname{f}_{hash}$ is label invariant,
    i.e.~$\operatorname{f}_{hash}(G^\sigma,\pi^\sigma) =
    \operatorname{f}_{hash}(G,\pi)$. By Lemma~\ref{lemma:R_correctness},
    the equality:
    $$\operatorname{f}_{hash}(G^\sigma,
    \bar{R}(G^\sigma,\pi_0^\sigma,[\nu]_i^\sigma)) =
    \operatorname{f}_{hash}(G^\sigma,\bar{R}(G,\pi_0,[\nu]_i)^\sigma) =
    \operatorname{f}_{hash}(G,\bar{R}(G,\pi_0,[\nu]_i))$$
    holds for each
    $i \le |\nu|$. Therefore, we have
    $\bar{\phi}(G^\sigma, \pi_0^\sigma,
    \nu^\sigma)=\bar{\phi}(G,\pi_0,\nu)$.
\end{proof}
  
\begin{replem}[\textbf{\ref{lemma:sound_rules}}]
 For each of the rules in our proof system, if all its premises are valid facts for the
 graph $(G,\pi_0)$, and all additional conditions required by the rule are fulfilled,
 then the facts that are derived by the rule are also valid facts for the graph $(G,\pi_0)$.
\end{replem}
 
\begin{proof}
  We prove the statement of the lemma for each of the rules of our
  proof system. For the rule \texttt{ColoringAxiom}, first notice that
  the fact $\pi_0 \xrightarrow{eq} \bar{R}(G,\pi_0,[\ ])$ is obviously
  valid, since $\bar{R}(G,\pi_0,[\ ])$ is obtained by invoking the
  procedure $\mathtt{make\_equitable}(G,\pi_0,\alpha)$, where $\alpha$
  is the set of all cells of $\pi_0$ (Algorithm~\ref{algo:refinement}). Also, the empty sequence is the root node of
  any search tree, so the fact $[\ ] \in \mathcal{T}(G,\pi_0)$ is also
  valid.
   
  For the rule \texttt{Individualize}, from the first premise we have
  that the coloring assigned to the node $\nu$ is equal to $\pi$. From
  the second premise, the sequence $[\nu, v]$ is also a node of the
  search tree $\mathcal{T}(G,\pi_0)$.  Since the coloring assigned to
  the node $[\nu,v]$ is constructed from $\pi$ by individualizing the
  vertex $v$ and then making the coloring equitable (Algorithm~\ref{algo:refinement}) the fact
  $\operatorname{individualize}(\pi,v) \xrightarrow{eq}
  \bar{R}(G,\pi_0,[\nu,v])$ is also valid.

  For the rule \texttt{SplitColoring}, from the premise we have that
  the coloring assigned to the node $\nu$ is obtained by invoking the
  procedure $\mathtt{make\_equitable}$ for $\pi$. A consequence of
  Lemma~\ref{lemma:make_equitable} is that the coloring obtained after
  the first iteration of the \verb|while| loop in the procedure
  $\mathtt{make\_equitable}$ (Algorithm~\ref{algo:calculate_coloring})
  will be the coloring $split(G,\pi,i)$, where
  $i = min\{j\ |\ split(G,\pi,j) \prec \pi\}$ (and the rule assumes
  that such $i$ exists). Therefore, the rest of the iterations of the
  \verb|while| loop will start with the coloring $split(G,\pi,i)$ and
  produce the coloring $\bar{R}(G,\pi_0,\nu)$ at the end of the
  loop. It is easy to see that the same result would be obtained if
  $\mathtt{make\_equitable}$ was invoked for the coloring
  $split(G, \pi, i)$, that is, we have
  $split(G, \pi, i) \xrightarrow{eq} \bar{R}(G,\pi_0,\nu)$.

  For the rule \texttt{Equitable}, the premise tells us that the
  coloring assigned to the node $\nu$ is obtained from $\pi$ by
  invoking the procedure $\mathtt{make\_equitable}$. Since the rule
  assumes that there are no cells for which
  $\operatorname{split}(G,\pi,i) \prec \pi$, the coloring $\pi$ is
  equitable, so the application of the procedure
  $\mathtt{make\_equitable}$ on $\pi$ produces $\pi$ itself. This
  means that $\bar{R}(G,\pi_0,\nu) = \pi$.

  For the rule \texttt{TargetCell}, its premise is the fact
  $\bar{R}(G,\pi_0, \nu) = \pi$, which claims that $\pi$ is the
  coloring assigned to the node $\nu$. Since the target cell is always
  the first non-singleton cell of the coloring assigned to a node (and
  the rule assumes that such a cell exists for the coloring $\pi$),
  the first fact the rule derives is valid.  Furthermore, the
  sequences of the form $[\nu, v]$, where $v$ belongs to the target
  cell corresponding to $\nu$ are also the nodes of the tree, so the
  remaining facts derived by the rule are also valid.

  For the rule \texttt{InvariantAxiom}, the fact
  $\bar{\phi}(G,\pi_0,[\ ]) = \bar{\phi}(G \pi_0,[\ ])$ is obviously
  true, by reflexivity.

  For the rule \texttt{InvariantsEqual}, if the invariants are equal
  for the parent nodes of the two given nodes $\nu_1$ and $\nu_2$
  (which is one of the premises of the rule), and $\operatorname{f}_{hash}$ function
  takes equal values on $(G,\pi_1)$ and $(G,\pi_2)$, where $\pi_1$ and
  $\pi_2$ are the colorings corresponding to $\nu_1$ and $\nu_2$
  respectively, then the node invariants are also equal for $\nu_1$
  and $\nu_2$.

  The correctness of the rule \texttt{InvariantsEqualSym} follows from
  the symmetry of the equality.

  For the rule \texttt{OrbitsAxiom}, the fact of the form
  $\{ v \} \vartriangleleft \operatorname{orbits}(G,\pi_0,\nu)$
  derived by the rule is always valid, provided that $v$ is a vertex
  of $G$ (which is assumed by the rule).

  For the rule \texttt{MergeOrbits}, let $\sigma \in Aut(G,\pi_0)$,
  $w_1 \in \Omega_1$ and $w_2\in \Omega_2$ satisfy the conditions
  required by the rule, and let $u_1$ and $u_2$ be arbitrarily chosen
  vertices from $\Omega_1$ and $\Omega_2$, respectively. Since we have
  the facts
  $\Omega_1 \vartriangleleft \operatorname{orbits}(G,\pi_0,\nu)$ and
  $\Omega_2 \vartriangleleft \operatorname{orbits}(G,\pi_0, \nu)$ as
  premises, it holds that $\Omega_1$ and $\Omega_2$ are indeed subsets
  of some orbits with respect to $Aut(G,\pi)$, where
  $\pi=\bar{R}(G,\pi_0,\nu)$. Therefore, there exist
  $\sigma_1,\sigma_2 \in Aut(G,\pi)$ such that $w_1^{\sigma_1}=u_1$
  and $w_2^{\sigma_2}=u_2$. Since $\nu^\sigma = \nu$, the permutation
  $\sigma$ also belongs to $Aut(G,\pi)$ (because
  $\pi^\sigma = \bar{R}(G,\pi_0,\nu)^\sigma =
  \bar{R}(G^\sigma,\pi_0^\sigma,\nu^\sigma) = \bar{R}(G,\pi_0,\nu) =
  \pi$).  Now we have $u_2 = u_1^{\sigma_1^{-1}\sigma\sigma_2}$, where
  $\sigma_1^{-1}\sigma\sigma_2 \in Aut(G,\pi)$, so $u_1$ and $u_2$
  belong to the same orbit with respect to $Aut(G,\pi)$. Therefore,
  $\Omega_1 \cup \Omega_2$ is a subset of some orbit with respect to
  $Aut(G,\pi)$.

  For \texttt{PruneInvariant}, from premises and the condition
  $\operatorname{f}_{hash}(G,\pi_1) > \operatorname{f}_{hash}(G,\pi_2)$ it
  follows
  $\bar{\phi}(G,\pi_0,[\nu',v']) > \bar{\phi}(G,\pi_0,[\nu'',v'')$,
  since the node invariants are compared lexicographically. The node
  invariant function $\bar{\phi}$ satisfies the axiom $\phi1$
  (Lemma~\ref{lemma:phi_correct}), so the invariant of any leaf
  $\nu_1$ of the tree $\mathcal{T}(G,\pi_0,[\nu',v'])$ is greater than
  the invariant of any leaf $\nu_2$ of the tree
  $\mathcal{T}(G,\pi_0,[\nu'',v''])$. Therefore, the canonical leaf
  cannot belong to the tree $\mathcal{T}(G,\pi_0,[\nu'',v''])$.
  
  For the rule \texttt{PruneLeaf}, we have two cases -- either $\pi_1$
  is not discrete, or $G^{\pi_1} > G^{\pi_2}$.  In the first case, the
  node $\nu'$ is not a leaf, and all the leaves from
  $\mathcal{T}(G,\pi_0,\nu')$ have greater node invariants than the
  node $\nu''$ (since
  $\bar{\phi}(G,\pi_0,\nu') = \bar{\phi}(G,\pi_0,\nu'')$ and node
  invariants are compared lexicographically). Therefore, $\nu''$ is
  not the canonical leaf. In the second case, the nodes $\nu'$ and
  $\nu''$ are both leaves with equal node invariants, but the graph
  that corresponds to the leaf $\nu'$ is greater than the graph
  corresponding to the leaf $\nu''$. Again, the leaf $\nu''$ is not
  the canonical leaf.
  
  For \texttt{PruneAutomorphism}, since $\sigma \in Aut(G,\pi_0)$ and
  $\nu'^\sigma = \nu''$, we have
  $\mathcal{T}(G,\pi_0,\nu'') =
  \mathcal{T}(G^\sigma,\pi_0^\sigma,\nu'^\sigma) =
  \mathcal{T}(G,\pi_0,\nu')^\sigma$, i.e.~the trees
  $\mathcal{T}(G,\pi_0,\nu')$ and $\mathcal{T}(G,\pi_0,\nu'')$
  correspond to each other with respect to $\sigma$. Assume the
  opposite, that $\mathcal{T}(G,\pi_0,\nu'')$ contains the canonical
  leaf $\nu^*$ of $\mathcal{T}(G,\pi_0)$. Now the leaf
  ${\nu^*}^{\sigma^{-1}}$ belongs to the tree
  $\mathcal{T}(G,\pi_0,\nu')$.  Let $\pi^* = \bar{R}(G,\pi_0,\nu^*)$
  and $\pi'= \bar{R}(G,\pi_0,{\nu^*}^{\sigma^{-1}})$.  Since $\sigma$
  is an automorphism of $(G,\pi_0)$, we have
  $\sigma^{-1}\pi'=\pi'^\sigma =
  \bar{R}(G,\pi_0,{\nu^*}^{\sigma^{-1}})^\sigma =
  \bar{R}(G,\pi_0,\nu^*)=\pi^*$, so $\sigma=\pi'{\pi^*}^{-1}$. Now it
  holds
  $(G,\pi_0) = (G,\pi_0)^\sigma = (G, \pi_0)^{\pi'{\pi^*}^{-1}}=((G,
  \pi_0)^{\pi'})^{{\pi^*}^{-1}}$, and thus it follows
  $(G,\pi_0)^{\pi'} = (G,\pi_0)^{\pi^*} = C(G,\pi_0)$.  Therefore, the
  leaf ${\nu^*}^{\sigma^{-1}}$ also corresponds to the canonical form
  $C(G,\pi_0)$.  But ${\nu^*}^{\sigma^{-1}}$ is lexicographically
  smaller than $\nu^*$, which contradicts the assumption that $\nu^*$
  is the canonical leaf.
  
  For the rule \texttt{PruneOrbits}, since $w_1$ and $w_2$ belong to the
  same orbit $\Omega$ with respect to $Aut(G,\pi)$ (where
  $\pi = \bar{R}(G,\pi_0,\nu)$), there exists $\sigma \in Aut(G,\pi)$
  such that $w_2 = w_1^\sigma$. Since $\pi^\sigma = \pi$, and
  $\pi \prec \pi_0$, it also holds $\pi_0^\sigma = \pi_0$, so
  $\sigma \in Aut(G,\pi_0)$. Moreover, for any vertex $v$ it holds
  $\pi(v^\sigma) = \pi^\sigma(v^\sigma) = \pi(v)$, so if $v \in \nu$,
  it must be $v^\sigma = v$, since $v$ is in a singleton cell of
  $\pi$. Therefore, $[\nu,w_1]^\sigma = [\nu,w_2]$, and
  $\mathcal{T}(G,\pi_0,[\nu,w_2]) =
  \mathcal{T}(G,\pi_0,[\nu,w_1]^\sigma) =
  \mathcal{T}(G^\sigma,\pi_0^\sigma, [\nu,w_1]^\sigma) =
  \mathcal{T}(G,\pi_0,[\nu,w_1])^\sigma$, so subtrees
  $\mathcal{T}(G,\pi_0,[\nu,w_1])$ and
  $\mathcal{T}(G,\pi_0,[\nu,w_2])$ correspond to each other with
  respect to $\sigma$. Similarly as for the rule
  \texttt{PruneAutomorphism}, we can now prove that the canonical leaf
  cannot belong to $\mathcal{T}(G,\pi_0,[\nu,w_2])$, since
  $[\nu,w_1] < [\nu,w_2]$.
  
  For the rule \texttt{PruneParent}, we may notice that if all the
  children of a node $\nu$ are not ancestors of the canonical leaf,
  then $\nu$ is also not an ancestor of the canonical leaf.

  For the rule \texttt{PathAxiom}, the derived fact
  $\operatorname{on\_path}(G,\pi_0,[\ ])$ is obviously valid, since
  the canonical leaf always exists, so the root node must be its
  ancestor.

  For the rule \texttt{ExtendPath}, from its first premise it holds
  that $\nu$ is an ancestor of the canonical leaf. By the second
  premise of the rule, the children of the node $\nu$ are the nodes
  $[\nu, w']$, where $w' \in W$, and the rule assumes that $w$ is one
  of them. According to the rest of the premises, all of the children
  of $\nu$, except the node $[\nu, w]$, are not ancestors of the
  canonical leaf. This means that the node $[\nu, w]$ must be an
  ancestor of the canonical leaf.

  Finally, for the rule \texttt{CanonicalLeaf}, by its first premise
  it holds that the node $\nu$ is an ancestor of the canonical
  leaf. Since the rule assumes that the coloring $\pi$ (assigned to
  the node $\nu$ by second premise) is discrete, the node $\nu$ is a
  leaf, and therefore it \emph{is} the canonical leaf, so
  $(G^\pi,\pi_0^\pi)$ is the canonical form of the graph $(G,\pi_0)$.
\end{proof}

\begin{replem}[\textbf{\ref{lemma:valid_facts}}]
  For any proof that corresponds to the graph $(G,\pi_0)$, all the
  facts that it derives are valid facts for $(G,\pi_0)$.
\end{replem}

\begin{proof}
  The lemma can be proven by the total induction on the proof
  length. That is, we can assume that the statement holds for each fact
  that can be derived by a proof of a length smaller than $k$, and
  then prove that the statement also holds for each fact that is
  derived by a proof of the length exactly $k$. If a fact $F$ is
  derived by a proof of the length exactly $k$, we can assume that $F$
  is obtained by the last rule application in that proof (otherwise
  there would exist a shorter proof of the same fact).  A rule can be
  applied only if all its premises are already derived (that is, by
  proofs shorter than $k$, which means that they are valid, by the
  induction hypothesis), and all the additional conditions required by
  the rule are fulfilled. According to Lemma~\ref{lemma:sound_rules},
  the derived fact $F$ is then also valid. By proving the induction
  step, we have proven the lemma.
\end{proof}

\begin{repthm}[\textbf{\ref{thm:soundness}}]
  Let $(G,\pi_0)$ be a colored graph. Assume a proof that corresponds
  to $(G, \pi_0)$ such that it derives the fact
  $C(G,\pi_0)=(G',\pi')$. Then $(G', \pi')$ is the canonical form of
  the graph $(G,\pi_0)$.
\end{repthm}

\begin{proof}
  The theorem is a direct consequence of
  Lemma~\ref{lemma:valid_facts}, since all the facts that are derived
  by the proof are valid facts, including the fact
  $C(G,\pi_0)=(G',\pi')$, which means that $(G',\pi')$ is the
  canonical form of the graph $(G,\pi_0)$.
\end{proof}

\begin{replem}[\textbf{\ref{lemma:comp_R_T_node}}]
  Let $\nu$ be a node of the search tree $\mathcal{T}(G,\pi_0)$ and
  let $\pi$ be the coloring assigned to $\nu$. Then the facts
  $\nu \in \mathcal{T}(G,\pi_0)$ and $\bar{R}(G,\pi_0,\nu) = \pi$ can
  be derived in our proof system. Furthermore, if $\pi$ is not
  discrete, and $W$ is the first non-singleton cell of $\pi$, then the
  fact $\bar{T}(G,\pi_0,\nu)=W$ can also be derived.
\end{replem}

\begin{proof}
  First we prove that for any node $\nu$ from $\mathcal{T}(G,\pi_0)$
  and any coloring $\pi_1$ of $G$, if we assume the fact
  $\pi_1 \xrightarrow{eq} \bar{R}(G,\pi_0,\nu)$, then we can derive a
  fact $\bar{R}(G,\pi_0,\nu) = \pi_2$ for some equitable coloring
  $\pi_2$ such that $\pi_2 \preceq \pi_1$. We can prove this statement
  using the well-founded induction, since the relation $\prec$ is
  well-founded (we have finitely many distinct colorings of $G$). Let
  us assume that the statement holds for all colorings strictly finer
  than $\pi_1$, and let us prove that the statement then also holds
  for $\pi_1$. If $\pi_1$ is itself equitable, then we can apply the
  rule $\mathtt{Equitable}$ to derive the fact
  $\bar{R}(G,\pi_0,\nu) = \pi_1$, and $\pi_1 \preceq \pi_1$, so the
  statement also holds for $\pi_1$. Otherwise, let $i$ be the smallest
  color of $\pi_1$ such that
  $\operatorname{split}(G,\pi_1, i) \prec \pi_1$.  Using the rule
  $\mathtt{SplitColoring}$, we can derive the fact
  $\operatorname{split}(G,\pi_1,i) \xrightarrow{eq}
  \bar{R}(G,\pi_0,\nu)$.  Now, since
  $\operatorname{split}(G,\pi_1,i) \prec \pi_1$, by inductive
  hypothesis there is an equitable coloring $\pi_2$ finer than
  $\operatorname{split}(G,\pi_1,i)$ such that we can derive the fact
  $\bar{R}(G,\pi_0,\nu) = \pi_2$, assuming the fact
  $\operatorname{split}(G,\pi_1, i) \xrightarrow{eq}
  \bar{R}(G,\pi_0,\nu)$. But the fact
  $\operatorname{split}(G,\pi_1, i) \xrightarrow{eq}
  \bar{R}(G,\pi_0,\nu)$ is derived from the fact
  $\pi_1 \xrightarrow{eq} \bar{R}(G,\pi_0,\nu)$ by one application of
  the $\mathtt{SplitColoring}$ rule, so we can also derive the fact
  $\bar{R}(G,\pi_0,\nu) = \pi_2$ assuming the fact
  $\pi_1 \xrightarrow{eq} \bar{R}(G,\pi_0,\nu)$. Therefore, the
  statement also holds for $\pi_1$. By well-founded induction, the
  statement now holds for all colorings of $G$.
  
  Now we can prove the main statement of the lemma, by induction on
  $k=|\nu|$. For $k=0$, we have $\nu = [\ ]$. The fact
  $[\ ] \in \mathcal{T}(G,\pi_0)$ can be derived by applying the rule
  $\mathtt{ColoringAxiom}$. The same rule also derives the fact
  $\pi_0 \xrightarrow{eq} \bar{R}(G,\pi_0,[\ ])$.  By the previous
  statement, there is an equitable coloring $\pi_0'$ finer than
  $\pi_0$ such that we can derive the fact
  $\bar{R}(G,\pi_0,[\ ]) = \pi_0'$. By Lemma~\ref{lemma:valid_facts},
  $\pi_0'$ must be the coloring assigned to the node $[\ ]$. Finally,
  if $\pi_0'$ is not discrete, we can apply the rule
  $\mathtt{TargetCell}$ to derive the fact $\bar{T}(G,\pi_0,[\ ]) = W$
  from the fact $\bar{R}(G,\pi_0,[\ ]) = \pi_0'$, where $W$ is the
  first non-singleton cell of $\pi_0'$. Therefore, we have proven that
  the lemma holds for $k=0$.

  Let us now assume that the lemma holds for some $k$, and let $\nu$
  be an arbitrary node of $\mathcal{T}(G,\pi_0)$ such that
  $|\nu|=k+1$. If $\nu = [\nu', v]$, then $\nu'$ is also a node of
  $\mathcal{T}(G,\pi_0)$, and $v$ must belong to $W'$, where $W'$ is
  the target cell corresponding to $\nu'$. Since $|\nu'|=k$, by the
  induction hypothesis, we can derive the facts
  $\nu' \in \mathcal{T}(G,\pi_0)$, $\bar{R}(G,\pi_0,\nu') = \pi'$,
  where $\pi'$ is the coloring assigned to the node $\nu'$, and
  $\bar{T}(G,\pi_0,\nu')=W'$.  Moreover, the fact
  $\bar{T}(G,\pi_0,\nu')=W'$ can be only derived by the rule
  \texttt{TargetCell}, and this rule also derives the fact
  $[\nu', v] \in \mathcal{T}(G,\pi_0)$, since $v \in W'$.  Now, by
  application of the rule $\mathtt{Individualize}$, we can derive the
  fact
  $\operatorname{individualize}(\pi', v) \xrightarrow{eq} \bar{R}(G,\pi_0,[\nu',
  v])$. By the previous statement, there exists an equitable coloring
  $\pi''$ finer than $\operatorname{individualize}(\pi',v)$ such that we can
  derive the fact $\bar{R}(G,\pi_0,[\nu',v]) = \pi''$, assuming the
  fact
  $\operatorname{individualize}(\pi', v) \xrightarrow{eq}
  \bar{R}(G,\pi_0,[\nu',v])$ (that we have already derived). By
  Lemma~\ref{lemma:valid_facts}, the coloring $\pi''$ must be the
  coloring assigned to the node $[\nu',v]$. Again, if $\pi''$ is not
  discrete, by applying the rule $\mathtt{TargetCell}$ we can derive
  the fact $\bar{T}(G,\pi_0,[\nu',v]) = W$, where $W$ is the first
  non-singleton cell of $\pi''$. Therefore, we have proven that the
  lemma also holds for $k+1$. By proving the induction step, we have
  proven the lemma.
\end{proof}

\begin{replem}[\textbf{\ref{lemma:comp_phi}}]
  Let $\nu_1$ and $\nu_2$ be two nodes such that $|\nu_1|=|\nu_2|$,
  and that have equal node invariants.  Then there is a proof that
  derives the fact
  $\bar{\phi}(G,\pi_0,\nu_1)=\bar{\phi}(G,\pi_0,\nu_2)$.
\end{replem}

\begin{proof}
  We prove this lemma by induction on $k = |\nu_1|=|\nu_2|$. For
  $k = 0$, we have $\nu_1 = \nu_2 = [\ ]$, and since there is a proof
  that derives the fact $[\ ] \in \mathcal{T}(G,\pi_0)$
  (Lemma~\ref{lemma:comp_R_T_node}), we can apply the rule
  $\mathtt{InvariantAxiom}$ to derive the fact
  $\bar{\phi}(G,\pi_0,[\ ]) = \bar{\phi}(G,\pi_0,[\ ])$.  Assume that
  the statement holds for some $k$, and let $\nu_1 =[\nu',v']$ and
  $\nu_2 =[\nu'',v'']$ be arbitrary nodes such that
  $|\nu'|=|\nu''|=k$. Let $\pi_1$ and $\pi_2$ be the colorings that
  are assigned to nodes $\nu_1$ and $\nu_2$, respectively. By
  Lemma~\ref{lemma:comp_R_T_node}, there are proofs of the facts
  $\bar{R}(G,\pi_0,\nu_1) = \pi_1$ and
  $\bar{R}(G,\pi_0,\nu_2) = \pi_2$. Since the node invariants of
  $\nu_1$ and $\nu_2$ are equal, and since the node invariants are
  compared lexicographically, it follows that the node invariants of
  their parent nodes $\nu'$ and $\nu''$ are also equal, and that
  $\operatorname{f}_{hash}(G,\pi_1)=\operatorname{f}_{hash}(G,\pi_2)$. By the
  induction hypothesis, there is a proof of the fact
  $\bar{\phi}(G,\pi_0,\nu')=\bar{\phi}(G,\pi_0,\nu'')$. Now we can
  apply the rule $\mathtt{InvariantsEqual}$ to derive the fact
  $\bar{\phi}(G,\pi_0,\nu_1) = \bar{\phi}(G,\pi_0,\nu_2)$. By proving
  the induction step, we have proven the lemma.
\end{proof}

\begin{replem}[\textbf{\ref{lemma:comp_leaf}}]
  If $\nu$ is a non-canonical leaf of $\mathcal{T}(G,\pi_0)$, then
  there is a proof that derives the fact
  $\operatorname{pruned}(G,\pi_0,[\nu]_i)$ for some $i \le |\nu|$.
\end{replem}

\begin{proof}
  Let $\nu^*$ be the canonical leaf of $\mathcal{T}(G,\pi_0)$, and let
  $|\nu^*| = l$.  Let $\nu$ be a non-canonical leaf, and let
  $|\nu|=k$. Since $\nu$ is not canonical, one of the following cases
  must be true:
  \begin{itemize}
  \item $\bar{\phi}(G,\pi_0,\nu^*) > \bar{\phi}(G,\pi_0,\nu)$: since
    the node invariants are compared lexicographically, we have the
    following two subcases:
    \begin{itemize}
    \item there is some $m$, such that $m \le k$ and $m \le l$, and
      such that
      $\bar{\phi}(G,\pi_0,[\nu^*]_m) >
      \bar{\phi}(G,\pi_0,[\nu]_m)$. Assume that $m$ is the smallest
      number with that property. Let $\pi_1$ and $\pi_2$ be the
      colorings assigned to the nodes $[\nu^*]_m$ and $[\nu]_m$.  Now
      we have that $\bar{\phi}(G,\pi_0, [\nu^*]_{m-1})$ and
      $\bar{\phi}(G,\pi_0,[\nu]_{m-1})$ are equal, and
      $\operatorname{f}_{hash}(G,\pi_1) >
      \operatorname{f}_{hash}(G,\pi_2)$. Since there are proofs that
      derive the fact
      $\bar{\phi}(G,\pi_0, [\nu^*]_{m-1}) =
      \bar{\phi}(G,\pi_0,[\nu]_{m-1})$ (Lemma~\ref{lemma:comp_phi}),
      and the facts $\bar{R}(G,\pi_0,[\nu^*]_m) = \pi_1$ and
      $\bar{R}(G,\pi_0,[\nu]_m) = \pi_2$ (Lemma~\ref{lemma:comp_R_T_node}),
      we can apply the rule $\mathtt{PruneInvariant}$ to derive the
      fact $\operatorname{pruned}(G,\pi_0,[\nu]_m)$.
    \item the invariant $\bar{\phi}(G,\pi_0,\nu)$ is a prefix of
      $\bar{\phi}(G,\pi_0,\nu^*)$. This means that $k < l$, and the
      invariants $\bar{\phi}(G,\pi_0,[\nu^*]_k)$ and
      $\bar{\phi}(G,\pi_0,\nu)$ are equal. Let $\pi_1$ and $\pi_2$ be
      the colorings assigned to the nodes $[\nu^*]_k$ and $\nu$,
      respectively. Since $\nu$ is a leaf, the coloring $\pi_2$ is
      discrete. On the other hand, the node $[\nu^*]_k$ is not a leaf,
      so $\pi_1$ is not discrete. By Lemma~\ref{lemma:comp_phi}, there
      is a proof that derives the fact
      $\bar{\phi}(G,\pi_0,[\nu^*]_k) = \bar{\phi}(G,\pi_0,\nu)$, and
      by Lemma~\ref{lemma:comp_R_T_node} there are proofs that derive the
      facts $\bar{R}(G,\pi_0,[\nu^*]_k)=\pi_1$ and
      $\bar{R}(G,\pi_0,\nu) = \pi_2$. Thus, we can apply the rule
      $\mathtt{PruneLeaf}$ to derive the fact
      $\operatorname{pruned}(G,\pi_0,\nu)$.
      \end{itemize}
    \item $\bar{\phi}(G,\pi_0,\nu^*) = \bar{\phi}(G,\pi_0,\nu)$, but
      $G_{\nu^*} > G_{\nu}$: since the node invariants of $\nu^*$ and
      $\nu$ are equal, it must hold $l=k$, because node invariants are
      compared lexicographically. By Lemma~\ref{lemma:comp_phi}, there
      is a proof that derives the fact
      $\bar{\phi}(G,\pi_0,\nu^*) = \bar{\phi}(G,\pi_0,\nu)$. Let
      $\pi^*$ and $\pi$ be the colorings assigned to $\nu^*$ and
      $\nu$, respectively. By Lemma~\ref{lemma:comp_R_T_node}, there are
      proofs deriving the facts $\bar{R}(G,\pi_0, \nu^*) =\pi^*$ and
      $\bar{R}(G,\pi_0, \nu)=\pi$. Since $\nu^*$ and $\nu$ are leaves,
      both $\pi^*$ and $\pi$ are discrete. From
      $G_{\nu^*} = G^{\pi^*}$ and $G_\nu=G^\pi$ we have
      $G^{\pi^*}>G^{\pi}$. Now we can apply the rule
      $\mathtt{PruneLeaf}$ to derive the fact
      $\operatorname{pruned}(G,\pi_0,\nu)$.
    \item $\bar{\phi}(G,\pi_0,\nu^*) = \bar{\phi}(G,\pi_0,\nu)$ and
      $G_{\nu^*} = G_{\nu}$, but $\nu^* <_{lex} \nu$: again, since the
      invariants of $\nu^*$ and $\nu$ are equal, it must hold $l =
      k$. Assume that $\pi^*=\bar{R}(G,\pi_0,\nu^*)$ and
      $\pi=\bar{R}(G,\pi_0,\nu)$.  Notice that $G^{\pi^*} = G^\pi$
      implies $(G^{\pi^*},\pi_0^{\pi^*}) = (G^\pi,\pi_0^\pi)$, since
      both $\pi^*$ and $\pi$ are discrete colorings finer than
      $\pi_0$. Now it holds $(G,\pi_0)^\sigma=(G,\pi_0)$, for
      $\sigma = \pi^*\pi^{-1}$, so $\sigma \in Aut(G,\pi_0)$. Since
      ${\pi^*}^\sigma=\sigma^{-1}\pi^*=(\pi^*\pi^{-1})^{-1}\pi^*=\pi{\pi^*}^{-1}\pi^*=\pi$,
      we have
      $\bar{R}(G,\pi_0,\nu) = \pi = {\pi^*}^\sigma =
      \bar{R}(G,\pi_0,\nu^*)^\sigma=\bar{R}(G^\sigma,\pi_0^\sigma,{\nu^*}^\sigma)
      = \bar{R}(G,\pi_0, {\nu^*}^\sigma)$, so it must hold
      ${\nu^*}^\sigma = \nu$, since no two distinct leaves correspond
      to the same coloring. Finally, by
      Lemma~\ref{lemma:comp_R_T_node}, there are proofs of the facts
      $\nu^* \in \mathcal{T}(G,\pi_0)$ and
      $\nu \in \mathcal{T}(G,\pi_0)$. Now we can apply the rule
      $\mathtt{PruneAutomorphism}$ to derive the fact
      $\operatorname{pruned}(G,\pi_0,\nu)$.\footnote{In fact, if $i$
        is the smallest number such that $[\nu^*]_i <_{lex} [\nu]_i$,
        we can derive the fact
        $\operatorname{pruned}(G,\pi_0,[\nu]_i)$.}
    \end{itemize}
    Thus, in all cases, we can derive the fact
    $\operatorname{pruned}(G,\pi_0,[\nu]_i)$ for some $i \le |\nu|$.
\end{proof}

\begin{replem}[\textbf{\ref{lemma:comp_tree}}]
  If a subtree $\mathcal{T}(G,\pi_0,\nu)$ does not contain the
  canonical leaf, then there is a proof that derives the fact
  $\operatorname{pruned}(G,\pi_0,[\nu]_i)$ for some $i \le |\nu|$.
\end{replem}

\begin{proof}
  Let define the \emph{height} $h(\nu)$ of a node
  $\nu \in \mathcal{T}(G,\pi_0)$ as a maximal distance to some of its
  descendant leaves.  We prove the lemma by induction on the node
  height. If $h(\nu)=0$, then $\nu$ is a leaf, and the statement
  follows from Lemma~\ref{lemma:comp_leaf}. Let us assume that the
  statement holds for all nodes whose height is smaller than some
  $l>0$, and let $\nu$ be an arbitrary node such that
  $h(\nu)=l$. Assume that $W$ is a target cell assigned to the node
  $\nu$. Then, by Lemma~\ref{lemma:comp_R_T_node}, there is a proof of
  the fact $\bar{T}(G,\pi_0,\nu)=W$. For each $w \in W$, the height of
  the node $[\nu,w]$ is smaller than $l$, and the subtree
  $\mathcal{T}(G,\pi_0,[\nu,w])$ also does not contain the canonical
  leaf, so by induction hypothesis, there is a proof that derives
  either the fact $\operatorname{pruned}(G,\pi_0,[\nu,w])$, or a fact
  $\operatorname{pruned}(G,\pi_0,[\nu]_i)$ for some $i \le |\nu|$. If
  the second case is true for at least one $w \in W$, then the
  statement also holds for the node $\nu$. Otherwise, for each
  $w\in W$ there is a proof of the fact
  $\operatorname{pruned}(G,\pi_0,[\nu,w])$, so we can apply the rule
  $\mathtt{PruneParent}$ to derive the fact
  $\operatorname{pruned}(G,\pi_0,\nu)$. By proving the induction step,
  we proved the lemma.
\end{proof}

\begin{replem}[\textbf{\ref{lemma:comp_path}}]
  If a node $\nu$ is an ancestor of the canonical leaf of the tree
  $\mathcal{T}(G,\pi_0)$, then there is a proof that derives the fact
  $\operatorname{on\_path}(G,\pi_0,\nu)$.
\end{replem}

\begin{proof}
  The lemma will be proved by induction on $k = |\nu|$. For $k = 0$,
  the node $\nu$ is the root node $[\ ]$. We can apply the rule
  $\mathtt{PathAxiom}$ to derive the fact
  $\operatorname{on\_path}(G,\pi_0, [\ ])$. Assume that the statement
  holds for some $k$, and let $\nu = [\nu',v]$ be an ancestor of the
  canonical node such that $|\nu'|=k$. Since $\nu'$ is also an
  ancestor of the canonical leaf, by induction hypothesis there is a
  proof that derives the fact
  $\operatorname{on\_path}(G,\pi_0,\nu')$. Let $W$ be the target cell
  assigned to the node $\nu'$. By Lemma~\ref{lemma:comp_R_T_node},
  there is a proof that derives the fact $\bar{T}(G,\pi_0,\nu') =
  W$. Notice that $v \in W$ and for each $w \in W \setminus \{ v \}$
  the node $[\nu',w]$ is not an ancestor of the canonical leaf, that
  is, the subtree $\mathcal{T}(G,\pi_0,[\nu',w])$ does not contain the
  canonical leaf. By Lemma~\ref{lemma:comp_tree}, either there is a
  proof that derives the fact
  $\operatorname{pruned}(G,\pi_0,[\nu',w])$, or a proof deriving the
  fact $\operatorname{pruned}(G,\pi_0,[\nu']_i)$ for some $i \le
  k$. But in the second case the pruned node $[\nu']_i$ would be an
  ancestor of the canonical leaf, which is not possible, by
  Lemma~\ref{lemma:valid_facts}. Therefore, there exists a proof that
  derives the fact $\operatorname{pruned}(G,\pi_0,[\nu',w])$ for each
  $w \in W \setminus \{ v \}$. Thus, we can apply the rule
  $\mathtt{ExtendPath}$ to derive the fact
  $\operatorname{on\_path}(G,\pi_0,[\nu',v])$. We proved the induction
  step, so the lemma holds.
\end{proof}

\begin{replem}[\textbf{\ref{lemma:comp_canon}}]
  If $\pi^*$ is the coloring that corresponds to the canonical leaf
  $\nu^*$ of the tree $\mathcal{T}(G,\pi_0)$, then there is a proof
  that derives the fact $C(G,\pi_0) = (G^{\pi^*},\pi_0^{\pi^*})$.
\end{replem}

\begin{proof}
  Since the canonical leaf $\nu^*$ is an ancestor of itself, by
  Lemma~\ref{lemma:comp_path}, there is a proof that derives the fact
  $\operatorname{on\_path}(G,\pi_0,\nu^*)$. Also, by
  Lemma~\ref{lemma:comp_R_T_node}, there is a proof that derives the
  fact $\bar{R}(G,\pi_0,\nu^*) = \pi^*$. Since $\nu^*$ is a leaf,
  $\pi^*$ is discrete, so we can apply the rule
  $\mathtt{CanonicalLeaf}$ to derive the fact
  $C(G,\pi_0) = (G^{\pi^*}, \pi_0^{\pi^*})$.
\end{proof}

\begin{repthm}[\textbf{\ref{thm:completeness}}]
  Let $(G,\pi_0)$ be a colored graph, and let $(G',\pi')$ be its
  canonical form.  Then there is a proof that corresponds to
  $(G,\pi_0)$ deriving the fact $C(G,\pi_0)=(G',\pi')$.
\end{repthm}

\begin{proof}
  By Lemma~\ref{lemma:comp_canon}, we can derive the fact
  $C(G,\pi_0)=(G^{\pi^*}, \pi_0^{\pi^*})$, where $\pi^*$ is the
  coloring corresponding to the canonical leaf $\nu^*$. Since the
  colored graph $(G^{\pi^*}, \pi_0^{\pi^*})$ is the canonical graph of
  $(G,\pi_0)$ by definition, our theorem holds.
\end{proof}

\end{document}